\documentclass[aps,pra,twocolumn,showpacs,superscriptaddress]{revtex4}
\usepackage{graphicx}
\begin{document}

\title{Quantum transport of strongly interacting photons in a one-dimensional nonlinear waveguide}

\author{Mohammad Hafezi \cite{JQI-address}}
 \affiliation{Physics Department, Harvard University, Cambridge, MA - 02138}
\author{Darrick E. Chang}
\affiliation{Center for the Physics of Information and Institute for Quantum Information,
California Institute of Technology, Pasadena, CA 91125}
\author{Vladimir Gritsev}
\affiliation{Physics Department, University of Fribourg, Chemin du Musee 3, 1700 Fribourg, Switzerland}
\author{Eugene Demler}
\affiliation{Physics Department, Harvard University, Cambridge, MA - 02138}
\author{Mikhail D. Lukin}
\affiliation{Physics Department, Harvard University, Cambridge, MA - 02138}

\date{\today}
\begin{abstract}
We present a theoretical technique for solving the quantum
transport problem of a few photons through a one-dimensional,
strongly nonlinear waveguide. We specifically consider the
situation where the evolution of the optical field is governed by
the quantum nonlinear Schr\"odinger equation (NLSE). Although this
kind of nonlinearity is quite general, we focus on a realistic
implementation involving cold atoms loaded in a hollow-core
optical fiber, where the atomic system provides a tunable
nonlinearity that can be large even at a single-photon level. In
particular, we show that when the interaction between photons is
effectively repulsive, the transmission of multi-photon components
of the field is suppressed. This leads to anti-bunching of the
transmitted light and indicates that the system acts as a
single-photon switch. On the other hand, in the case of attractive
interaction, the system can exhibit either anti-bunching or
bunching, which is in stark contrast to semiclassical
calculations. We show that the bunching behavior is related to the
resonant excitation of bound states of photons inside the system.
\end{abstract} \pacs{42.65.-k,05.60.Gg,42.50.-p}
\maketitle
\section{Introduction}

Physical systems that enable single photons to interact strongly
with each other are extremely valuable for many emerging
applications. Such systems are expected to facilitate the
construction of single-photon switches and transistors
\cite{Kimble:2005,Rempe2008,Chang:2007}, networks for quantum
information processing, the realization of strongly correlated
quantum systems using light
\cite{Hartmann:2006,Greentree:2006,Chang:2008} and the investigation of novel new many-body physics such as out of equilibrium behaviors. One potential
approach involves the use of high-finesse optical microcavities
containing a small number of resonant atoms that mediate the
interaction between photons \cite{Haroche:RMP,Kimble:2005}. Their
nonlinear properties are relatively straightforward to analyze or
simulate because they involve very few degrees of freedom
(\textit{i.e.}, a single optical mode)
\cite{Imamoglu:1997,Grangier:1998,Imamoglu:1998}. Recently, an
alternative approach has been suggested, involving the use of an
ensemble of atoms coupled to propagating photons in
one-dimensional, tightly-confining optical waveguides
\cite{ghosh:093902,Kien:PRA2008,Akimov:2008}. Here, the
nonlinearities are enhanced due to the transverse confinement of
photons near the diffraction limit and the subsequent increase in
the atom-photon interaction strength. The propagation of an
optical field inside such a nonlinear medium (\textit{e.g.},
systems obeying the quantum nonlinear Schr\"odinger equation) is
expected to yield much richer effects than the case of an optical
cavity due to the large number of spatial degrees of freedom
available. Simultaneously, however, these degrees of freedom make
analysis much more difficult and in part cause these systems to
remain relatively unexplored
\cite{Lai:1989,kartner93,drummond:lecture,ShanhuiFan:PRL2007,Chang:2008}.
We show that the multi-mode, quantum nature of the system plays an
important role and results in phenomena that have no analogue in
either single-mode cavities or classical nonlinear optics. It is
interesting to note that similar low-dimensional, strongly
interacting condensed matter systems are an active area of
research, but most of this work is focused on closed systems close
to the ground state or in thermal equilibrium
\cite{lieb-liniger,korepin:1993,Kinoshita:2004,Parades:2004,Caux:PRL2007}.
On the other hand, as will be seen here, the relevant regime for
photons often involves open systems and driven dynamics.

In this article, we develop a technique to study the quantum
transport of a few photons inside a finite-length, strongly
nonlinear waveguide where the light propagation is governed by the
quantum nonlinear Schr\"odinger equation~(NLSE), and apply this
technique to study the operation of this system as a single-photon
switch. In particular, we study the transmission and reflection
properties of multi-photon fields from the system as well as
higher-order correlation functions of these fields. We find that
these correlations not only reflect the switching behavior, but
reveal some aspects of the rich structure associated with the
spatial degrees of freedom inside the system, which allow photons
to {}``organize'' themselves. In the regime where an effectively
repulsive interaction between photons is achieved, anti-bunching
in the transmitted field is observed because of the switching
effect, and is further reinforced by the tendency of photons to
repel each other. In the attractive regime, either anti-bunching
(due to switching) or bunching can occur. We show that the latter
phenomenon is a clear signature of the creation of photonic bound
states in the medium. Although we focus on a particular
realization involving the propagation of light, our conclusions on
quantum transport properties are quite general and valid for any
bosonic system obeying the NLSE.

This article is organized as follows. In Sec.~\ref{sec:Model}, we
describe an atomic system whose interactions with an optical field
can be manipulated using quantum optical techniques such that the
light propagation obeys the quantum NLSE. This method relies upon
electromagnetically induced transparency~(EIT) to achieve
resonantly enhanced optical nonlinearities with low propagation
losses and the trapping of stationary light pulses using spatially
modulated control fields. Before treating the nonlinear properties
of the system, we first consider the linear case in
Sec.~\ref{sec:linear}, where it is shown that the light trapping
technique leads to a field build-up inside the medium and a set of
discrete transmission resonances, much like an optical cavity. In
Sec.~\ref{sec:semi-classical}, we then investigate the nonlinear
transport properties of the system such as reflectivity and
transmittivity in the semi-classical limit, where the NLSE is
treated as a simple complex differential equation. Here we find
that the presence of the nonlinearity causes the transmission
resonances to shift in an intensity-dependent way -- the system
behaves as a low-power, nonlinear optical switch, whose behavior
does not depend on the sign of the nonlinear interaction. In
Sec.~\ref{sec:quantum_formalism}, we present a full quantum
formalism to treat the NLSE transport problem in the few-photon
limit. Sec.~\ref{sec:Analytical} is dedicated to analytical
solutions of the NLSE with open boundary conditions when the
system is not driven. In particular, we generalize the Bethe
ansatz technique to find the resonant modes of the system, which
help to elucidate the dynamics in the case of the driven system.
The driven system is studied in Sec.~\ref{sec:Results}, where
numerical solutions are presented along with a detailed study of
the different regimes of behavior. In particular, we find that the
correlation functions for the transmitted light do depend on the
sign of the nonlinear interaction, in contrast to what the
semi-classical calculations would suggest. We conclude in
Sec.~\ref{sec:Conclusion}.

\section{Model:  Photonic NLSE in 1D waveguide\label{sec:Model}}

In this section, we consider the propagation of light inside an
finite-length atomic medium under EIT conditions and with a Kerr
nonlinearity. We also describe a technique that allows for these
pulses of light to be trapped within the medium using an effective
Bragg grating formed by additional counter-propagating optical
control fields. We show that in the limit of large optical depth
the evolution of the system can be described by a nonlinear
Schr\"odinger equation.

\begin{figure}

\includegraphics[width=0.40\textwidth]{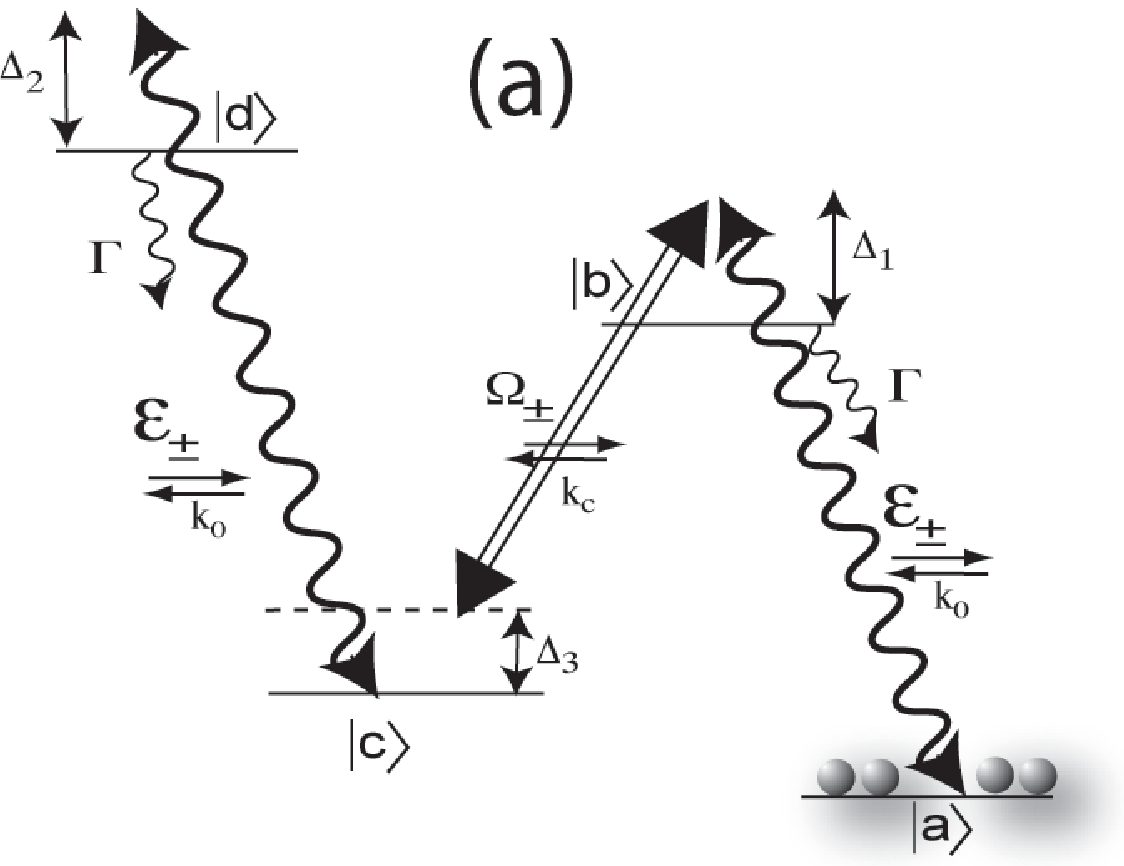} 
\includegraphics[width=0.40\textwidth]{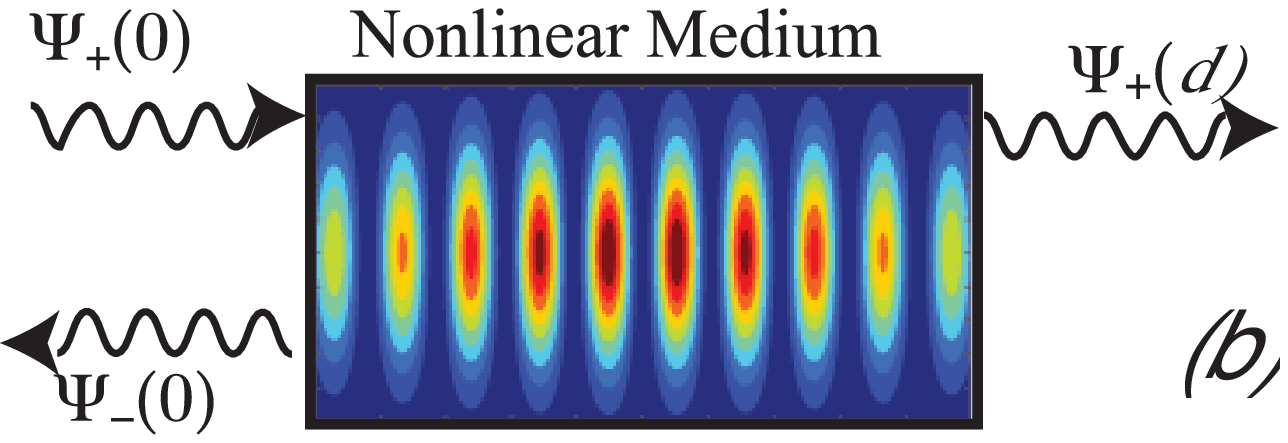}

\caption [Four-level atomic system for creating strong nonlinearity] {(a) Four-level atomic system for creating strong nonlinearity. Counter-propagating
control fields modulate the EIT for the forward- and backward-propagating
probe, and $|c\rangle\rightarrow|d\rangle$ transition gives rise
to a Kerr-type nonlinearity. (b) The light is confined
in the transverse direction due to the presence of the waveguide and
experiences an effective Bragg grating due to the presence of the
counter-propagating light.}

\label{fig:four_level}
\end{figure}

Following Ref.~\cite{Chang:2008}, we consider an ensemble of atoms
with the four-level internal structure shown in Fig.
\ref{fig:four_level}, which interact with counter-propagating
quantum fields with slowly-varying envelopes
$\hat{\mathcal{E}}_{\pm}$ inside an optical waveguide. These
fields are coupled to a spin coherence between states $|a\rangle$
and $|c\rangle$ via two classical, counter-propagating control
fields with Rabi frequencies $\Omega_{\pm}$ largely detuned from
the $|b\rangle\rightarrow|c\rangle$ transition. The case where the
fields propagate only in one direction~(say in the ``+''
direction) and where the detuning is zero corresponds to the usual
EIT system, where the atomic medium becomes transparent to
$\hat{\mathcal{E}}_{+}$ and the group velocity can be dramatically
slowed due to coupling between the light and spin wave~(so-called
``dark-state polaritons'') \cite{Fleischhauer:2000kx}. On the other hand, the presence of
counter-propagating control fields creates an effective Bragg
grating that causes the fields $\hat{\mathcal{E}}_{\pm}$ to
scatter into each other. This can modify the photonic density of
states and create a bandgap for the quantum fields. This photonic
bandgap prevents a pulse of light from propagating and can be used
to effectively trap the light inside the
waveguide~\cite{Andre:bandgap,Bajcsy:2003}. The trapping
phenomenon is crucial because it increases the time over which
photons can interact inside the medium. The presence of an
additional, far-detuned transition $|c\rangle\rightarrow|d\rangle$
that is coupled to $\hat{\mathcal{E}}_{\pm}$ leads to an
intensity-dependent energy shift of level $|c\rangle$, which
translates into a Kerr-type optical nonlinearity
\cite{Schmidt:1996}.

We now derive the evolution equations for the quantum fields. We
assume that all atoms are initially in their ground states
$|a\rangle$. To describe the quantum properties of the atomic
polarization, we define collective, slowly-varying atomic
operators, averaged over small but macroscopic volumes containing
$N_{z}\gg1$ particles at position $z$, \begin{equation}
\hat{\sigma}_{\alpha\beta}(z,t)=\frac{1}{N_{z}}\sum_{i=1}^{N_{z}}|\alpha_{i}\rangle\langle\beta_{i}|.\end{equation}
The collective atomic operators obey the following commutation
relations,

\begin{equation}
[\hat{\sigma}_{\alpha\beta}(z),\hat{\sigma}_{\mu\nu}(z')]=\frac{L}{N}\delta(z-z')(\delta_{\beta\mu}\hat{\sigma}_{\alpha\nu}(z)-\delta_{\alpha\nu}\hat{\sigma}_{\mu\beta}(z)),\end{equation}
while the forward and backward quantized probe fields in the ${\rm
{z}}$ direction obey bosonic commutation relations (at equal
time),
\begin{equation}
[\mathcal{\hat{E}}_{+}(z),\mathcal{\hat{E}}_{+}^{\dagger}(z')]=\delta(z-z').\end{equation}
The Hamiltonian for this system in the rotating frame can be
written as

\begin{eqnarray}
\hat{H} = & -&\frac{N}{L}\int [\hat{\sigma}_{bc}(z)(\Omega_{+}e^{ik_{c}z}+\Omega_{-}e^{-ik_{c}z})+H.c. ]\\ 
 & + & g\sqrt{2\pi}[\hat{\sigma}_{ba}(z)(\mathcal{\hat{E}}_{+}e^{ik_{0}z}+\mathcal{\hat{E}}_{-}e^{-ik_{0}z})+H.c.]\nonumber\\
 & + & g\sqrt{2\pi}[ \hat{\sigma}_{dc}(z)(\mathcal{\hat{E}}_{+}e^{ik_{0}z}+\mathcal{\hat{E}}_{-}e^{-ik_{0}z})+H.c.]\nonumber\\
& + &  \Delta_{1}\hat{\sigma}_{bb}(z)+\Delta_{3}\hat{\sigma}_{cc}(z)+(\Delta_{2}+\Delta_{3})\hat{\sigma}_{dd}(z) dz\nonumber\end{eqnarray}
where $g=\mu\sqrt{\frac{\omega_{ab}}{4\pi\hbar\epsilon_{0}A}}$ is
the atom-field coupling strength, $\mu$ is the atomic dipole
matrix element, and $A$ is the effective area of the waveguide
modes. For simplicity, we have assumed that the transitions $a$-$b$
and $c$-$d$ have identical coupling strengths $g$ and have ignored
transverse variation in the fields. The terms $\Delta_{i}$ denote
the light field-atomic transition detunings as shown in
Fig.~\ref{fig:four_level}(a). $k_c$ is the wavevector of the
control fields, while $k_0=n_{b}\omega_{ab}/c$ characterizes the
fast-varying component of the quantum field and $n_b$ is the
background refractive index. We also define $n_{0}=N/L$ as the
linear density of atoms in the ${\rm {z}}$ direction, and
$v_g=c\Omega^{2}/2{\pi}g^{2}n_0$ as the group velocity that the
quantum fields would have if they were not trapped by the Bragg
grating~(we will specifically be interested in the situation where
$\Omega_{+}=\Omega_{-}=\Omega$). Following
Ref.~\cite{Fleischhauer:2000kx}, we can define dark-state
polariton operators that describe the collective excitation of
field and atomic spin wave, which in the limit of slow group
velocity $\eta=\frac{c}{2v_{g}}\gg1$ are given by
$\hat{\Psi}_{\pm}=\frac{g\sqrt{2\pi
n_0}}{\Omega_{\pm}}\mathcal{\hat{E}}_{\pm}$. These operators obey
bosonic commutation relations
$[\hat{\Psi}_{\pm}(z),\hat{\Psi}_{\pm}^{\dagger}(z')]=\delta(z-z')$.
The definition of the polariton operators specifies that the
photon flux entering the system at its boundary is equal to the
rate that polaritons are created at the boundary inside the system
-- \textit{i.e.},
$c\langle\mathcal{\hat{E}_{+}}^{\dagger}\mathcal{\hat{E}_{+}}\rangle=v_{g}\langle\hat{\Psi}_{+}^{\dagger}\hat{\Psi}_{+}\rangle$.
In other words, excitations enter~(and leave) the system as
photons with velocity $c$, but inside the waveguide they are
immediately converted into polariton excitations with group
velocity $v_g$. Field correlations will also be mapped in a
similar fashion -- in particular, correlation functions that we
calculate for polaritons at the end of the waveguide $z=L$ will
also hold for the photons transmitted from the system. The total
number of polaritons in the system is given by
$\mathcal{N}_{pol.}=\int \langle\hat{\Psi}_{+}^{\dagger}\hat{\Psi}_{+}\rangle+\langle\hat{\Psi}_{-}^{\dagger}\hat{\Psi}_{-}\rangle dz$. 
The optical fields coupled to the atomic coherences of both the
$a-b$ and $c-d$ transitions are governed by Maxwell-Bloch
evolution equations,
\begin{equation}
\left(\frac{\partial}{\partial t}\pm c\frac{\partial}{\partial
z}\right)\mathcal{\hat{E}}_{\pm}(z,t)=ig\sqrt{2\pi}n_{0}(\hat{\sigma}_{ab}+\hat{\sigma}_{cd}).\label{eq:maxwellbloch}\end{equation}

Similar to the photonic operators, the atomic coherences can also
be written in terms of slowly-varying components,
\begin{eqnarray}
\hat{\sigma}_{ab} & = & \hat{\sigma}_{ab}^{+}e^{ik_{0}z}+\hat{\sigma}_{ab}^{-}e^{-ik_{0}z},\\
\hat{\sigma}_{cd} & = &
\hat{\sigma}_{cd}^{+}e^{ik_{0}z}+\hat{\sigma}_{cd}^{-}e^{-ik_{0}z}.\end{eqnarray}
We note that higher spatial orders of the coherence are thus
neglected. In practice, these higher orders are destroyed due to
atomic motion and collisions as atoms travel distances greater
than an optical wavelength during the typical time of the
experiment~\cite{Zimmer:2006}. Alternatively, one can use dual-V
atomic systems that do not require this approximation
\cite{Zimmer:2008}.

In the weak excitation limit~$(\hat{\sigma}_{aa}\simeq1)$, the
population in the excited state $|b\rangle$ can be neglected,
$\langle\hat{\sigma}_{bb}\rangle{\approx}0$. In this limit, the
evolution of the atomic coherence is given by
\begin{eqnarray} 
\dot{\hat{\sigma}}_{ab}^{\pm} & = &
(i\Delta_{1}-\Gamma/2)\hat{\sigma}_{ab}^{\pm}\\
&+& ig\sqrt{2\pi}\mathcal{\hat{E}}_{\pm}+i\Omega_{\pm}\hat{\sigma}_{ac}e^{\pm
i\Delta kz}\nonumber \end{eqnarray} where $\Delta k=k_{c}-k_{0}$ and
$\Gamma$ is the total spontaneous emission rate of state $b$~(for
simplicity we also assume that state $d$ has an equal spontaneous
emission rate). In the adiabatic limit where
$g\sqrt{2\pi}\langle\hat{\sigma}_{ac}\mathcal{\hat{E}}_{\pm}\rangle\ll\Gamma$,
the coherence $\hat{\sigma}_{ad}$  can be approximated by

\begin{equation}
\hat{\sigma}_{ad}\simeq\frac{g\sqrt{2\pi}\hat{\sigma}_{ac}}{-\Delta_{2}-\Delta_{3}-i\frac{\Gamma}{2}}(\mathcal{\hat{E}}_{+}e^{+ik_{0}z}+\mathcal{\hat{E}}_{-}e^{-ik_{0}z}).\end{equation}
Therefore, the spin wave evolution can be written as,
\begin{eqnarray}
\dot{\hat{\sigma}}_{ac} & = & i\Delta_{3}\hat{\sigma}_{ac}+ i(\hat{\sigma}_{ab}^{+}\Omega_{+}^{*}e^{-i\Delta kz}+\hat{\sigma}_{ab}^{-}\Omega_{-}^{*}e^{i\Delta kz})\nonumber\\
&+&\frac{2\pi ig^{2}}{-\Delta_{2}-\Delta_{3}-i\frac{\Gamma}{2}}(\mathcal{\hat{E}}_{+}^{\dagger}\mathcal{\hat{E}}_{+}+\mathcal{\hat{E}}_{-}^{\dagger}\mathcal{\hat{E}}_{-})\hat{\sigma}_{ac}
\end{eqnarray}

We now consider the situation where
$\Omega_{+}=\Omega_{-}=\Omega$, such that the counter-propagating
control fields form a standing wave. In the adiabatic limit
\cite{Fleischhauer:2000kx}, and keeping all terms up to third
order in the quantum fields, substituting these results into
Eq.~(\ref{eq:maxwellbloch}) and simplifying yields the following
evolution equations for the dark-state polariton operators,
\begin{eqnarray}
(c\partial_{z}+\partial_{t})\hat{\Psi}_{+}&=&-\frac{\xi}{2}(\hat{\Psi}_{+}-\hat{\Psi}_{-})-\frac{\eta}{2}\partial_{t}(\hat{\Psi}_{+}+\hat{\Psi}_{-})\nonumber\\
&-&i\Delta_{n}[(\hat{\Psi}_{+}^{\dagger}\hat{\Psi}_{+}+\hat{\Psi}_{-}^{\dagger}\hat{\Psi}_{-})(\hat{\Psi}_{+}+\hat{\Psi}_{-})\nonumber\\
&+&(\hat{\Psi}_{+}^{\dagger}+\hat{\Psi}_{-}^{\dagger})(\hat{\Psi}_{+}+\hat{\Psi}_{-})\hat{\Psi}_{+}] \\
(-c\partial_{z}+\partial_{t})\hat{\Psi}_{-}&=&+\frac{\xi}{2}(\hat{\Psi}_{+}-\hat{\Psi}_{-})-\frac{\eta}{2}\partial_{t}(\hat{\Psi}_{+}+\hat{\Psi}_{-})\nonumber\\
&-&i\Delta_{n}[(\hat{\Psi}_{+}^{\dagger}\hat{\Psi}_{+}+\hat{\Psi}_{-}^{\dagger}\hat{\Psi}_{-})(\hat{\Psi}_{+}+\hat{\Psi}_{-})\nonumber\\
&+&(\hat{\Psi}_{+}^{\dagger}+\hat{\Psi}_{-}^{\dagger})(\hat{\Psi}_{+}+\hat{\Psi}_{-})\hat{\Psi}_{-}],\label{eq:coupled_mode}\end{eqnarray}
where the linear dispersion is characterized by $\xi=\frac{2\pi
g^{2}n_0}{-i\Delta_{1}+\Gamma/2}$ . The nonlinearity coefficient
is given by the single photon AC-Stark shift:
$\Delta_{n}=\frac{\pi g^{2}}{2(\Delta_{2}+i\Gamma/2)}$. We note
that the wave-vector mismatch $\Delta k$ has been compensated for
by a small extra two-photon detuning equal to $(-\Delta kc/\eta)$.

The above equations describe the evolution of two coupled modes.
It is convenient to re-write these equations in terms of the
anti-symmetric and symmetric combinations
$A=(\Psi_{+}-\Psi_{-})/\sqrt{2}$ and
$S=(\Psi_{+}+\Psi_{-})/\sqrt{2}$. For large optical depths, we
then find that the anti-symmetric mode adiabatically follows the
symmetric mode, $A\simeq-(c/\xi)\partial_{z}S$. In this limit, the
evolution of the whole system can be described by a single
nonlinear Schr\"odinger equation,

\begin{equation}
\eta\frac{\partial}{\partial
t}S-\frac{c^{2}}{\xi}\frac{\partial^{2}}{\partial
z^{2}}S+8i\Delta_{n}S^{\dagger}S^{2}=0.\end{equation}

Physically, the coupling between $\hat{\Psi}_{\pm}$ induced by the
Bragg grating causes them to no longer behave independently, much
like the two counter-propagating components of an optical cavity
mode. We can write the above equation in dimensionless units by
introducing a characteristic length scale $L_{coh}=c/|{\rm
{Im}[\xi]|=c(\Delta_{1}^{2}+\Gamma{}^{2}/4)/2\pi
g^{2}n_{0}|\Delta_{1}|}$ and time scale $t_{coh}=\eta/|{\rm
{Im}[\xi]|=(\Delta_{1}^{2}+\Gamma{}^{2}/4)/2\Omega^{2}|\Delta_{1}|}$.
$L_{coh}$ corresponds to the length over which the field acquires
a $\pi$-phase in the propagation. The dimensionless NLSE then
reads

\begin{equation}
i\frac{\partial\tilde{S}}{\partial\tau}=-\frac{1}{2m}\frac{\partial^{2}\tilde{S}}{\partial\tilde{z}^{2}}+2\kappa\tilde{S}^{\dagger}\tilde{S}^{2},\label{eq:NLSE}\end{equation}
where for $\Delta_{1}<0$, the effective mass is
$m=\frac{1}{2}(1+i\frac{\Gamma}{2|\Delta_{1}|})$ and the
nonlinearity coefficient is
$\kappa=\frac{2\Delta_{n}}{c}=\frac{\pi
g^{2}/c}{\Delta_{2}+i\Gamma/2}$. Note that
$\tilde{\Psi}_{\pm}(z,t)$ and $\tilde{S}(z,t)$ are also in units
of $\sqrt{L_{coh}^{-1}}$, such that
$[\tilde{S}_{+}(\tilde{z}),\tilde{S}_{+}^{\dagger}(\tilde{z}')]=\delta(\tilde{z}-\tilde{z}')$.
For simplicity, we omit tilde superscripts in the following. We
can also write the nonlinear coefficient as
$\kappa=\frac{\Gamma_{1D}}{4(\Delta_{2}+i\Gamma/2)}$, where we
have identified $\Gamma_{1D}=4\pi g^{2}/c$ as the spontaneous
emission rate into the guided modes~($\Gamma_{1D}{\leq}\Gamma$). We are primarily interested
in the limit  $|\Delta_{1,2}|{\gg}\Gamma$ such that $m,\kappa$ are
mostly real and the evolution is dispersive.
Note that in this notation, the anti-symmetric combination of
forward and backward polaritons is given by
$A\simeq-i/2m\partial_{z}S\simeq-i\partial_{z}S$.

\section{Linear case: Stationary light enhancement \label{sec:linear}}

In this section, we investigate the linear transmission properties
of the signal field as a function of its frequency. The control
field leads to a Bragg grating that couples the forward and
backward components of the signal field together. We show that the
system therefore acts as an effective cavity whose \emph{finesse}
is determined by the optical density of the atomic medium.

\begin{figure}
\centering
\includegraphics[width=0.45\textwidth]{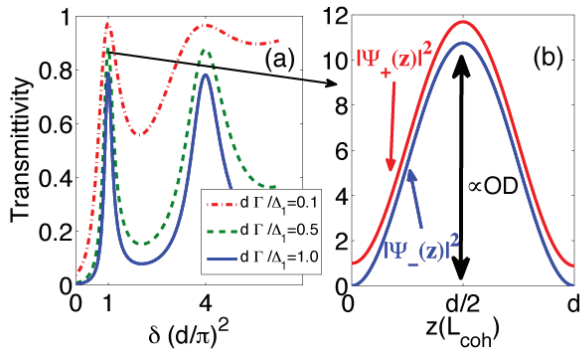}

\caption[Linear transmission spectrum ]{Linear case: (a) Transmittivtiy as a function of two-photon detuning.
Transmission peaks are attenuated because of linear loss on $|a\rangle\rightarrow|b\rangle$
transition which are plotted for three different loss rates $\beta=d~\Gamma/\Delta_{1}$.
(b) When the system is tuned on a transmission resonance ($\sqrt{\delta_{0}}d=n\pi$),
the field inside the medium is amplified.}

\label{fig:linear}
\end{figure}

For the linear case ($\kappa=0$), it is sufficient to treat the
forward and backward field operators as two complex numbers. In
the slow light regime ($\eta\gg1$), the coupled mode
equations~(Eqs.~\ref{eq:coupled_mode}) can be written in the Fourier
domain, with our dimensionless units, as
\begin{eqnarray}
\partial_{z}\Phi_{+} & = & \frac{i}{2}\delta(\Phi_{+}+\Phi_{-})+im(\Phi_{+}-\Phi_{-})\\
-\partial_{z}\Phi_{-} & = &
\frac{i}{2}\delta(\Phi_{+}+\Phi_{-})-im(\Phi_{+}-\Phi_{-}),\end{eqnarray}
where $\Psi_{+}(z,\tau)=\Phi_{+}(z,\delta)e^{-i\delta\tau}$ and
$\Psi_{-}(z,\tau)=\Phi_{-}(z,\delta)e^{-i\delta\tau}$and $\delta$
is the dimensionless two-photon detuning
$\delta=\Delta_{3}t_{coh}$. We specify that a classical field
$\Phi_{+}(z=0,\delta)=\alpha$ enters the system at $z=0$ with no
input at the other end of the system~($z=d$),
$\Phi_{-}(z=d,\delta)=0$, as shown in
~Fig. \ref{fig:four_level}(b). We note that $d=L/L_{coh}$ is the
length of the system in units of the coherence length introduced
earlier. For negligible losses~($|\Delta_{1}|{\gg}\Gamma$ ) and
$\Delta_{1}<0$, $m\simeq1/2$ and the profile of forward-going
polaritons inside the system will look like:

\begin{equation}
\frac{\Phi_{+}(z,\delta)}{\alpha}=\frac{2i\sqrt{\delta}\cos[(d-z)\sqrt{\delta}]+(1+\delta)\sin[(d-z)\sqrt{\delta}]}{2i\sqrt{\delta}\cos[d\sqrt{\delta}]+(1+\delta)\sin[d\sqrt{\delta}]}.\end{equation}
 Therefore, for a system with fixed length $d$, the transmission coefficient
varies with the frequency of the incident field, with transmission
resonances occurring at the values $\sqrt{\delta_{0}}d=n\pi$~($n$
is an integer). At these resonances, the system transmittance is
equal to one ($|\Phi(d,\delta)|=|\Phi(0,\delta)|$) and a field
build-up occurs inside the medium with a bell-shaped profile,
similar to a cavity mode~(see Fig.\ \ref{fig:linear}). The
positions of these resonances~(quadratic in $n$) reflect the
quadratic dispersion in Eq.~(\ref{eq:NLSE}). Note that in real
units, the positions of the resonances will depend on the
amplitude of the control field, since
$\Delta_{3}=\delta\frac{|\Delta_{1}|}{2|\Omega|^{2}}$. In the
limit of a coherent optically large system ($d\gg1$), the
intensity amplification in the middle of the system is equal to
$(d/2\pi)^{2}$ for the first resonance. In other words, the Bragg
scattering creates a cavity with an effective finesse proportional
to the square of the coherent length of the system
($\mathcal{F}\propto d^{2}$).

\begin{figure}
\centering
\includegraphics[width=0.4\textwidth]{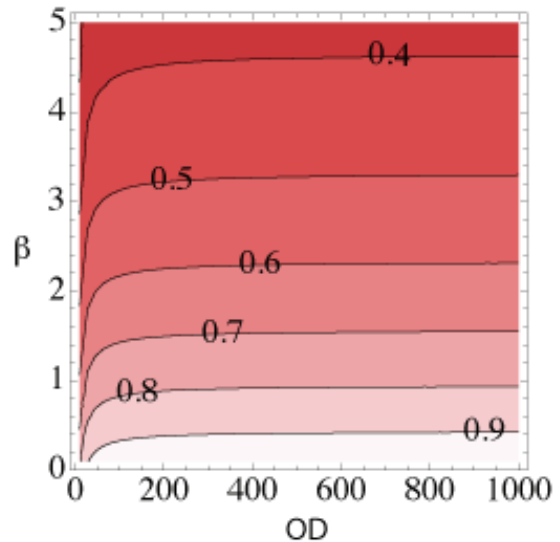}
\caption[Optimization of the transmission for different optical densities]{For large optical densities, the transmission on resonance ($\delta=\delta_0$), only depends on $\beta=OD \left(\frac{\Gamma}{\Delta_1|}\right)^2$. }

\label{fig:trans_beta_OD}
\end{figure}

We now derive the width of the first transmission resonance. For
small variations $\delta_{0}\pm\delta_b$ around the resonance
frequency, we can write
\begin{equation}
\frac{\Phi_{+}(d)}{\Phi_{+}(0)}=-1-\frac{i\pi}{4\delta_{0}^{3/2}}\delta_b+\frac{\pi^{2}}{16\delta_{0}^{3}}\delta_b^{2}+O(\delta_b^{3}).\end{equation}
Therefore, the width of the resonances~(say where it drops by
half) is given by
\begin{equation}
2\delta_b\simeq\delta_{0}^{3/2}=(\frac{\pi}{d})^{3}.\label{eq:width}\end{equation}
We have kept terms up to second order in $\delta_b$, since the
first order term does not give a decreasing correction to the
transmittance. While we have previously ignored absorption~(as
determined by the real part of $\xi$), we can estimate that its
effect is to attenuate the probe beam transmission by a factor
$\beta=d~\Gamma/|\Delta_{1}|$. As it is shown in Fig.\ \ref{fig:trans_beta_OD}, for large optical densities, $\beta$ can fully characterize the transmission coefficient on resonance $(\delta=\delta_0$). In particular, for a fixed $\beta$,  the resonant transmission is constant for any large optical density.  In other words, since the optical depth of the
system is given by $OD=d\frac{|\Delta_{1}|}{\Gamma}$, 
the transmittivity of the system remains constant for any $OD$
with the choice $|\Delta_{1}|=\Gamma\sqrt{OD/\beta}$.  In this case the effective cavity finesse for the system
becomes proportional to the optical density, \textit{i.e.},
$\mathcal{F}\propto OD$.

The total number of polaritons in the system can be estimated by,

\begin{eqnarray}
\mathcal{N}_{pol}&=&\int_{0}^{d}|\Phi_{+}(z)|^{2}+|\Phi_{-}(z)|^{2}dz\\
&=&\frac{(d^{2}+\pi^{2})^{2}}{4d\pi^{2}}|\Phi_{+}(0)|^{2}\simeq\frac{d^{3}}{4\pi^{2}}|\Phi_{+}(0)|^{2}.\nonumber\label{eq:number_polaritons}
\end{eqnarray}
This again shows that the polaritons experience many round trips
inside the system before exiting. In particular, if we define the
average intensity inside the medium as
$|\Phi_{+}^{ave}|^{2}=\mathcal{N}_{pol}/d$, then we readily
observe that the intensity of the polariton field is amplified
inside the medium by the square of the system size
$(|\Phi_{+}^{ave}|^{2}/|\Phi_{+}(0)|^{2}=(d/2\pi)^{2})$ --
\textit{i.e.} the finesse is proportional to the optical density
(OD).

The original proposal for observing an enhanced Kerr nonlinearity
with a four-level atomic system using EIT makes use of an optical
cavity to enhance the nonlinearity~\cite{Imamoglu:1997}. However,
as pointed out in Ref.~\cite{Grangier:1998}, the scheme suffers
from some inaccuracies in the effective Hamiltonian. More specifically, in Ref.
\cite{Imamoglu:1997}, the effective Hamiltonian was evaluated at
the center of the EIT transparency window. However, in practice,
EIT dramatically decreases the cavity linewidth because of the
large dispersion that accompanies the vanishing
absorption~\cite{Lukin:98}; this causes photons at frequencies
slightly shifted from the central frequency to be switched out of
the cavity. This leads to an extremely small allowable bandwidth
for the incoming photons~\cite{Grangier:1998} and was neglected in
the original analysis. We emphasize that the  analysis presented here takes into
account the dispersive properties of the medium, as we have
included the field dynamics up to second order in the detuning
from resonance~(this accounts for the effective mass of the
photons in our system). We verify the consistency of this
derivation in Appendix~\ref{sec:EIT-Bandgap} by solving the linear
system including full susceptibilities. It is  shown that
the results are consistent near the two-photon resonance (\textit{i.e.}, frequencies around $\delta=0$).

\section{Semi-classical nonlinear case\label{sec:semi-classical}}

\subsection{Dispersive regime\label{sub:semiclassical_dispersive}}

In this section, in contrast to the previous section, we include
the nonlinear term in the evolution equations to investigate its
effect in the semi-classical limit~(where the fields are still
treated as complex numbers). In this picture, the effect of
nonlinearity causes the transmission peaks to shift in frequency
in an intensity-dependent way to the left or right depending on
the sign of the nonlinearity coefficient $\kappa$. We show that
when $|\kappa| OD\gg1$, the magnitude of the shift is large even
at intensities corresponding to that of a single photon. In this
regime, we expect that the system can act as a single-photon
switch and that signatures of quantum transport will become
apparent~(the quantum treatment is described in
Sec.\ref{sec:quantum_formalism}).

Because of the self-phase modulation term in the evolution
equations~(Eqs.~\ref{eq:coupled_mode}), the forward and backward
fields acquire a phase shift proportional to their intensity.
Moreover, due to the conjugate-phase modulation terms, each field
undergo an extra phase shift proportional the intensity of the
other field. Classically, this yields a frequency shift in the
transmission spectrum when the nonlinearity is
small.
The shift in the transmission peak can be approximated by
$\Delta\delta\simeq2\kappa|\Phi_{+}^{ave}|^{2}$ where
$|\Phi_{+}^{ave}|^{2}\simeq\frac{d^{2}}{4\pi^{2}}|\Phi_{+}(0)|^{2}$
is the average intensity of polaritons in the system. Suppose that
we want the nonlinearity to be strong enough to shift the
transmission peaks at least by half of their widths,
$\Delta\delta\simeq\frac{1}{2}\delta^{3/2}$. Then, from
Eq.(\ref{eq:width}) this condition can be written as

\begin{equation}
|\Phi_{+}^{{\rm
{cr}}}(0)|^{2}=\left(\frac{\pi}{d}\right)^{5}\frac{1}{|\kappa|}\end{equation}

On the other hand, according to Eq.(\ref{eq:number_polaritons}),
we can write this condition in terms of the critical number of
polaritons inside the system,

\begin{equation}
\mathcal{N}_{pol}^{{\rm
{cr}}}=\frac{\pi^{3}}{4d^{2}\kappa}.\end{equation}

Since the nonlinearity coefficient is given by the light shift on
the $|c\rangle\rightarrow|d\rangle$ transition, in the dispersive
regime~($\Delta_{2}\gg\Gamma$), we have
$\kappa=\frac{1}{4}\frac{\Gamma_{1D}}{\Gamma}\frac{\Gamma}{\Delta_{2}}$.
Thus, we expect to have substantial nonlinearities at the level of
one polariton~(\textit{i.e.}, one incoming photon),
$\mathcal{N}_{pol.}^{{\rm {cr.}}}=1$, if

\begin{equation}
d^{2}=\pi^{3}\frac{\Gamma}{\Gamma_{1D}}\frac{\Delta_{2}}{\Gamma}\label{eq:singlephotonnonlinearity}\end{equation}
where $\Gamma_{1D}$ is the rate of spontaneous emission rate into
the guided modes. Strictly speaking, note that a single photon
cannot actually have a nonlinear phase shift~(as correctly derived
later using a fully quantum picture); however, we can still use
the results of this semiclassical calculation to qualitatively
understand the relevant physics.

We can also rewrite the above condition in term of the optical
density $(OD=d\frac{\Delta_{1}}{\Gamma})$ needed in the system.
From the linear case, we know that an optimal detuning, for a
transmission of 90\%, should satisfy
$d\frac{\Gamma}{\Delta_{1}}\sim0.5$. Then,
Eq.~(\ref{eq:singlephotonnonlinearity}) can be written as

\begin{equation}
OD=2\pi^{3}\frac{\Gamma}{\Gamma_{1D}}\frac{\Delta_{2}}{\Gamma}.\end{equation}
Taking for example a system where $\Delta_{2}\sim5\Gamma$ and
$\frac{\Gamma_{1D}}{\Gamma}\sim0.1$, nonlinearities at a
few-photon level can be observed for an optical density ${\rm
{OD}\simeq6200}$. 

\begin{figure}
\centering
\includegraphics[width=0.45\textwidth]{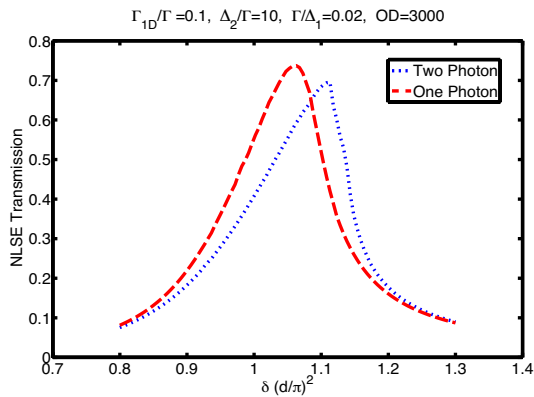}

\caption[Shifted resonances due to nonlinearity]{Due to nonlinearity, in the perturbative limit, the transmission spectrum
shifts for different intensities. The integrated intensities inside
the system is related to the number of present field quanta. OD=3000}

\label{fig:mode_shift}
\end{figure}

First, let us consider the case of positive $\kappa$. In Fig.
\ref{fig:mode_shift}, we observe that at large enough optical
density, the system can have very different transmission spectra
for \textit{low} and \textit{high} intensities that classically
correspond to having one and two polaritons (photons) inside the
system, respectively. Although we have ignored the quantization of
photons in this section, we can develop some insight into the
transmission properties of one- and two-photon states. Loosely
speaking, if we fix the input field frequency to lie at the
one-photon~(linear) transmission peak ($\delta_{0}$), the system
would block the transmission of incident two-photon states. More
realistically, suppose we drive the system with a weak classical
field~(coherent state), which can be well-approximated as
containing only zero, one, and two-photon components. We then
expect that the one-photon component will be transmitted through
the system, while the two-photon component will be reflected,
leading to anti-bunching of the transmitted light. We note that
the general spirit of this conclusion is sound; however, the
correct description of the system is achieved by taking into
account the quantization of photons which is presented in the next
sections.

\begin{figure}
\centering
\includegraphics[width=0.5\textwidth]{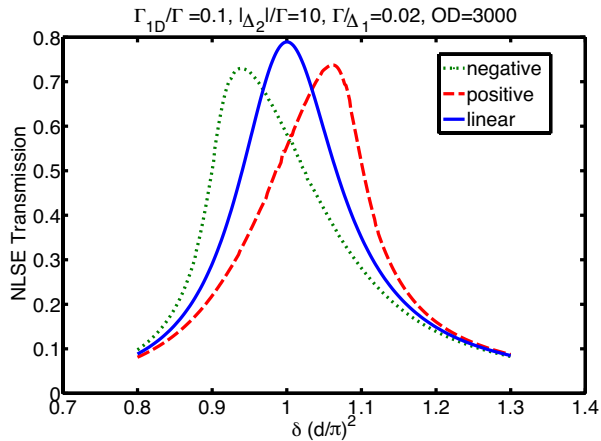}

\caption[Positive and negative nonlinearity in the semi-classical approximation]{For positive (negative) nonlinearity, in the perturbative limit, the
transmission spectrum shifts to the right (left) of linear transmission
spectrum (solid line), which is shown in dotted line (dashed line).
The incoming intensity corresponds to one-photon in the system.}

\label{fig:negative_vs_positive}
\end{figure}

A similar analysis holds for the case of negative $\kappa$. Note
that the sign of $\kappa$ depends on the detuning of the photonic
field from the atomic transition $|c\rangle\rightarrow|d\rangle$,
which can easily be adjusted in an experiment. This is in contrast
to conventional nonlinear optical fibers and nonlinear crystals,
where the nonlinearity coefficient is fixed both in magnitude and
sign. We find that a negative nonlinearity simply shifts the
transmission spectrum in the opposite direction as for the
positive case, as shown in Fig. \ref{fig:negative_vs_positive},
but all other conclusions remain the same. In particular, we would
expect anti-bunching to occur for this case as well, when a weak
coherent field is incident with its frequency fixed to the linear
transmission resonance. Surprisingly, the quantum
treatment~(Sec.\ref{sec:Results}), shows that the above conclusion
is wrong and system behaves very differently for negative
nonlinearity. We show that this difference in behavior can be
attributed to the presence of additional eigenstates~(photonic
bound states) in the medium and their excitation by the incident
field.

For even larger nonlinearities or intensities, the transmission
spectrum can become even more skewed and exhibit bistable
behavior, as similarly found in Ref.~\cite{Korsch:2008} in the
context of transport of Bose-Einstein condensates in one
dimension. There, the classical NLSE~(Gross-Pitaevskii equation)
was solved to find the mean-field transport properties of a
condensate scattering off a potential barrier.

Instead of considering the switching effect as a function of
number of photons inside the medium, we can also consider the
number of photons that need to be sent into the system. Clearly,
to have a well-defined transmission amplitude without substantial
pulse distortion, the incident pulse must be long enough so that
it fits within the bandwidth of the system resonance, as given in
Eq.~(\ref{eq:width}). To be specific, we consider an input pulse
whose duration is equal to the inverse of the bandwidth,
$t_{b}=(\frac{d}{\pi})^{3}t_{coh}$. We can relate the number of
incoming photons to an average incident intensity:

\begin{equation}
|\Phi_{+}(0)|^{2}=\mathcal{N}_{pol.}\frac{t_{b}}{t_{coh}}=\frac{\mathcal{N}_{pol.}}{(d/\pi)^{3}}\label{eq:ave-intensity}\end{equation}

\begin{figure}
\centering
\includegraphics[width=0.45\textwidth]{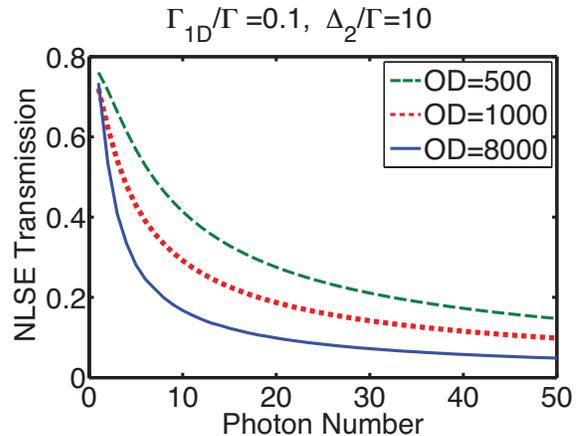}

\caption[Transmission versus photon number: dispersive case]{Transmission vs. the number of incident photons. For each
OD, $\Gamma/\Delta_{1}$ is chosen so that the system exhibit a
similar transmission for one photon.}

\label{fig:transmittivity_ph_number}
\end{figure}

Now, since the number of incident photons and
incoming polaritons are the same, we can assign an average amplitude to any incoming photons number  by Eq.\ (\ref{eq:ave-intensity}), and evaluate the transmission. Hence, we can
evaluate the number of incident photons needed to observe a
significant nonlinearity and saturate the system.
Fig. \ref{fig:transmittivity_ph_number} shows the transmittivity of
the nonlinear system as a function of number of photons in the
incoming wavepacket. We observe that for high optical densities
($OD>1000$), the transmittivity drops as the number of incoming
photons increases and the system gets saturated for even few
photons.

\subsection{Dissipative Regime}

In this section, we investigate the system in the presence of
nonlinear absorption, where $\kappa$ is imaginary. The nonlinear
dispersion of the previous case can simply be turned into
nonlinear absorption by setting the nonlinear detuning to zero
($\Delta_{2}=0,~\kappa=\frac{\Gamma_{1D}}{2i\Gamma}$). In the
quantum picture, this term does not affect the one-photon state,
while two-photon states can be absorbed by experiencing three
atomic transitions,
$|a\rangle\rightarrow|b\rangle\rightarrow|c\rightarrow|d\rangle$,
and subsequently being scattered from excited state $|d\rangle$ \cite{Harris:1998}. We consider the
quantum treatment of absorption later and first study the
semiclassical limit here.

\begin{figure}
\centering
\includegraphics[width=0.40\textwidth]{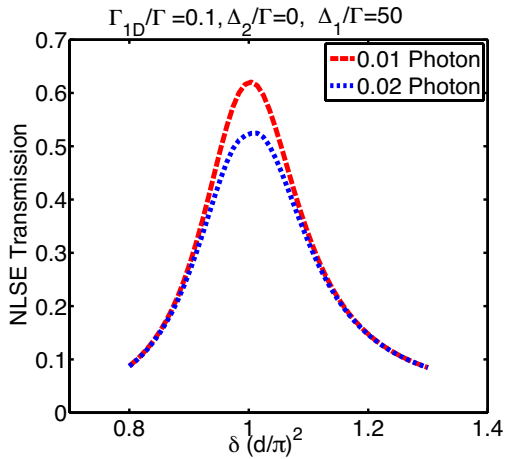}

\caption[Transmission versus photon number: absorptive case]{Due to nonlinear absorption, the transmission is
suppressed for higher intensities. The integrated intensities
inside the system is related to the number of present field
quanta. OD=3000}

\label{fig:transmittivity_abs}
\end{figure}

The presence of nonlinear absorption suppresses the transmission
of multi-photon states through the medium by causing them to
decay. This suppression becomes stronger for higher intensities as
shown in Fig. \ref{fig:transmittivity_abs}. We have used the same
optical density (OD) and 1D confinement ($\Gamma_{1D}/\Gamma$) as
in Fig. \ref{fig:mode_shift}. We observe that the effects of
nonlinear absorption are stronger than that of nonlinear
dispersion studied in Sec.\ref{sub:semiclassical_dispersive},
since it occurs at resonance~($\Delta_2=0$) where the atomic
response is strongest. It is thus possible to observe its effect
at even lower intensities, corresponding to effective photon
numbers two orders of magnitude smaller than the dispersive case.
Much like the dispersive case, the suppression of transmission of
multi-photon components should yield anti-bunching in the
transmitted field. In this case, however, these components are
simply lost from the system~(as opposed to showing up as a bunched
reflected field). 

\section{Quantum nonlinear formalism: Few-photon limit \label{sec:quantum_formalism}}

In this section, we describe a quantum mechanical approach that
enables one to solve the problem of quantum transport of a small
number of photons through the finite, nonlinear system described
in Sec.\ref{sec:Model}. This few-photon number limit is of
particular interest since it captures the physics of single-photon
switching.

We find it convenient to study the dynamics of the system of
photons in the Schr\"odinger picture, where one can explicitly solve
for the few-body wave functions. This approach is made possible by
truncating the Hilbert space so that only subspaces with $n_{max}$
photons are less are present. In the following, we will consider
the case where $n_{max}=2$, although our analysis can be easily
extended to cover any other value. This truncation is justified
when the incident coherent field is sufficiently weak that the
average photon number is much smaller than one inside the system
($|\alpha_{0}|^{2}d^{3}\ll1$, where $\alpha_{0}$ is the amplitude
of the incoming field). Thus, we can write the general state of
the system as:

\begin{eqnarray}
|\psi(t)\rangle&\simeq&\int
dz_{1}dz_{2}\phi(z_{1},z_{2},t)S^{\dagger}(z_{1})S^{\dagger}(z_{2})|0\rangle\\
&+&\int
dz\theta(z,t)S^{\dagger}(z)|0\rangle+\epsilon|0\rangle.\nonumber\end{eqnarray}
The first, the second and the third term correspond to two-photon,
one-photon and vacuum state, respectively. Note that because of
bosonic symmetrization, $\phi(z_{1},z_{2},t)$ should be symmetric
in $z_{1}$ and $z_{2}$. This formalism allows us to capture any
non-trivial spatial order between photons in our
system~(\textit{e.g.}, the de-localization of two photons as
represented by the off-diagonal terms in $\phi(z_{1},z_{2})$).
Since the NLSE Hamiltonian commutes with the field number operator
$\hat{S}^{\dagger}\hat{S}$, manifolds with different field quanta
are decoupled from each other inside the medium. Therefore, the
evolution for the one-photon and
two-photon manifolds under the NLSE Hamiltonian can be written as,\\
 \begin{eqnarray}
i\frac{\partial}{\partial t}\phi(z_{1},z_{2},t)&=&-\frac{1}{2m}\left(\frac{\partial^{2}}{\partial z_{1}^{2}}+\frac{\partial^{2}}{\partial z_{2}^{2}}\right)\phi(z_{1},z_{2},t)\nonumber\\
&+&2\kappa\phi(z_{1},z_{2},t)\delta(z_{1}-z_{2})\label{eq:two_photon}\end{eqnarray}

\begin{equation}
i\frac{\partial}{\partial
t}\theta(z,t)=-\frac{1}{2m}\frac{\partial^{2}}{\partial
z^{2}}\theta(z,t).\label{eq:one_photon}\end{equation}

However, the system is driven with an input field at $z=0$, which
allows different manifolds to be coupled at the boundaries. This
is analogous to fiber soliton experiments where a classical input
field mixes quantum solitons with different photon numbers
\cite{Drummond:Nature93,Haus:BOOK,Agrawal:2007}. 
In particular, for a classical input field,\\
 \begin{eqnarray}
\hat{\Psi}_{+}(z=0)|\psi(t)\rangle & = & \alpha(t)|\psi(t)\rangle\\
\hat{\Psi}_{-}(z=d)|\psi(t)\rangle & = &
0|\psi(t)\rangle,\end{eqnarray}
which corresponds to a coherent state with~(possibly
time-dependent) amplitude $\alpha(t)$ as an input at $z=0$, and no
input~(\textit{i.e.}, vacuum) at $z=d$. Since we specify that the
input coherent field is weak ($\alpha\ll1$), the amplitude of the
vacuum state is almost equal to one ($\epsilon\simeq1$). The annihilation
operator in these equations reduces the photon number on the
left-hand side by one. Thus, such boundary conditions relate
different photon subspaces whose photon number differ by one,
\textit{e.g.} the two-photon and one-photon wavefunctions. In the
adiabatic limit where the anti-symmetric part of the field
($A=(\hat{\Psi}_{+}-\hat{\Psi}_{-})/\sqrt{2}$) follows the
symmetric part ($S=(\hat{\Psi}_{+}+\hat{\Psi}_{-})/\sqrt{2}$),
we have\\
 \begin{equation}
\Psi_{+}=\frac{1}{\sqrt{2}}(S-\frac{i}{2m}\partial_{z}S)~~~,~~~\Psi_{-}=\frac{1}{\sqrt{2}}(S+\frac{i}{2m}\partial_{z}S).\end{equation}
Therefore the boundary conditions at $z=0$ can be re-written as
\begin{eqnarray} &\frac{1}{\sqrt{2}}&\int
dz_{1}dz_{2}~[S-\frac{i}{2m}\partial_{z}S]_{z=0}S^{\dagger}(z_{1})S^{\dagger}(z_{2})\phi(z_{1},z_{2},t)|0\rangle\nonumber\\
&=&\alpha\int
dz\theta(z,t)S^{\dagger}|0\rangle,\\
&\frac{1}{\sqrt{2}}&\int
dz~[S-\frac{i}{2m}\partial_{z}S]_{z=0}S^{\dagger}(z)\theta(z,t)|0\rangle=\alpha|0\rangle.\nonumber \end{eqnarray}
Using the identity
$\int[\partial_{z}S(z),S^{\dagger}(z')]f(z')dz'=\partial_{z}f(z)$, the boundary conditions on the
one-photon and two-photon wave functions
can be written as:\\
 \begin{eqnarray}
\phi(z_{1}=0,z_{2},t)-\frac{i}{2m}\partial^{(1)}\phi(z_{1},z_{2},t)|_{z_{1}=0}&=&\frac{\alpha}{\sqrt{2}}\theta(z_{2},t)\nonumber\\
\theta(z=0,t)-\frac{i}{2m}\partial_{z}\theta(z=0,t)&=&\sqrt{2}\alpha,\label{eq:boundary_condition}\end{eqnarray}
where $\partial^{(1)}$ acts on the first parameter. This type of
open boundary condition is known as a Robin or mixed boundary
condition, which involves a combination of both the function and
its derivative. In the present case, the open boundary conditions
allow particles to freely \emph{enter} and \emph{leave} the
system. We emphasize that this process is noise-less, in that the
loss of population from the interior of our system is related by
our boundary condition equations to the flow of particle current
through the system boundaries. This is in contrast to an optical
cavity, for instance, where photons inside the cavity leak
$\textit{dissipatively}$ into the environment \cite{Gardiner:1985}. Similarly the boundary condition at
$z=d$ reads
\begin{eqnarray}
\phi(d,z,t)+\frac{i}{2m}\partial^{(1)}\phi(d,z,t)&=&0\\
\theta(d,t)+\frac{i}{2m}\partial_{z}\theta(d,t)&=&0.\end{eqnarray}
Given the boundary conditions and the equations of motion in the
interior, we can completely solve for the photon wavefunctions.

Once the wavefunctions are determined, it is possible to determine
the intensity profile as well as any other correlation function
for the photons. For example, the intensity of the forward-going
polariton is

\begin{eqnarray}
I(z,t)&=&\langle\psi(t)|\hat{\Psi}_{+}^{\dagger}(z)\hat{\Psi}_{+}(z)|\psi(t)\rangle\\
&=&\langle1|\hat{\Psi}_{+}^{\dagger}(z)\hat{\Psi}_{+}(z)|1\rangle+\langle2|\hat{\Psi}_{+}^{\dagger}(z)\hat{\Psi}_{+}(z)|2\rangle,\nonumber\end{eqnarray}
where $|j\rangle$ denotes the component of the total wavefunction
$|{\psi(t)}\rangle$ containing $j$ photons. The first and second
terms on the right thus correspond to the one- and two-photon
contributions to the intensity. By re-writing expressions in terms
of $S$ instead of $(\hat{\Psi}_{+},\hat{\Psi}_{-})$, we obtain:

\begin{eqnarray}
I(z,t)&=&\frac{1}{2}\left|\theta(z)-\frac{i}{2m}\partial_{z}\theta(z)\right|^{2}\\
&+&2\int
dz'\left|\phi(z',z)-\frac{i}{2m}\partial^{(2)}\phi(z',z)\right|^{2}\nonumber\end{eqnarray}
Similarly, the second-order correlation function for the forward
field is
\begin{eqnarray}
&~&\langle\psi|\hat{\Psi}_{+}^{\dagger2}(z)\hat{\Psi}_{+}^{2}(z)|\psi\rangle = \label{eq:two_correlator}\\
&~&\left|\phi(z,z)-\frac{i}{m}\partial^{(1)}\phi(z,z) - \frac{1}{4m^{2}}\partial^{(1)}\partial^{(2)}\phi(z,z)\right|^{2} \nonumber \end{eqnarray}
which in our truncated space only depends on the two-photon
wave function. Now, we evaluate the normalized second-order
correlation function $g_{2}(z)$, which characterizes the photon
statistics of an arbitrary field. This function takes the form
\begin{equation}
g_{2}(z)=\frac{\langle\psi|\hat{\Psi}_{+}^{\dagger2}(z)\hat{\Psi}_{+}^{2}(z)|\psi\rangle}{\left|\langle\psi|\hat{\Psi}_{+}^{\dagger}(z)\hat{\Psi}_{+}(z)|\psi\rangle\right|^{2}},\end{equation}
and physically characterizes the relative probability of detecting
two consecutive photons at the same position $z$. If this quantity
is less (greater) than one, the photonic field is anti-bunched
(bunched). In particular, if $g_{2}(z)=0$, the field is perfectly
anti-bunched and there is no probability for two photons to
overlap in position. In our truncated Hilbert space, $g_{2}(z)$ of
the transmitted field is given by

\begin{equation}
g_{2}(z=d)\simeq\frac{\langle2|\Psi_{+}^{\dagger2}(d)\Psi_{+}^{2}(d)|2\rangle}{\left|\langle1|\Psi_{+}^{\dagger}(d)\Psi_{+}(d)|1\rangle+\langle2|\Psi_{+}^{\dagger}(d)\Psi_{+}(d)|2\rangle\right|^{2}}\label{eq:g2_0}\end{equation}
We note that this expression can be simplified, since at $z=d$, we
have $\Psi_{-}=\frac{1}{\sqrt{2}}(S+\frac{i}{2m}\partial_{z}S)=0$
and $\Psi_{+}=\sqrt{2}S$. Therefore,

\begin{equation}
g_{2}(d)=\frac{4|\phi(d,d)|^{2}}{\left(\left|\theta(d)\right|^{2}+4\int
dz'\left|\phi(z',d)\right|^{2}\right)^{2}}.\label{eq:g2_0_simple}\end{equation}
We can also evaluate the stationary two-time correlation, which is
defined in the Heisenberg picture as:

\begin{equation}
g_{2}(z,\tau)=\frac{\langle\psi|\Psi_{+}^{\dagger}(z,0)\Psi_{+}^{\dagger}(z,\tau)\Psi_{+}(z,\tau)\Psi_{+}(z,0)|\psi\rangle}{\left|\langle\psi|\Psi_{+}^{\dagger}(z,0)\Psi_{+}(z,0)|\psi\rangle\right|^{2}},\label{eq:g2tau}\end{equation}
where the denominator is simplified in the stationary steady-state
regime. This correlation function characterizes the probability of
detecting two photons at position $z$ but separated by time
$\tau$. We can re-write $g_{2}(z,\tau)$ in terms of wavefunctions
in the Schr\"odinger picture in the following way. We first note
that the expression
$|\tilde{\psi}(0)\rangle=\hat{\Psi}_{+}(z,0)|\psi\rangle$
appearing in the equation above can be thought of as a new
wavefunction, which describes the state of the system after a
photon is initially detected at time $t=0$ and position $z$. This
new state naturally has one less photon than the original state,
and by simplifying the expressions, it can be written as:

\begin{equation}
|\tilde{\psi}(0)\rangle=\int\theta^{{\rm {new}}}(z')S^{\dagger}(z')|0\rangle+\epsilon^{{\rm {new}}}|0\rangle\end{equation}
where the new one-photon and vacuum amplitudes are given by\\
 \begin{eqnarray}
\frac{\theta^{{\rm {new}}}}{\sqrt{2}} & = & \phi(d,z',t=0)-\frac{i}{2m}\partial^{(1)}\phi(d,z',t=0)\\
\epsilon^{{\rm {new}}} & = &
\frac{1}{\sqrt{2}}\left(\theta(d,t=0)-i\partial\theta(d,t=0)\right).\end{eqnarray}
Here we have assumed that $z=d$, since we are interested in the
transmitted field. Now, Eq.~(\ref{eq:g2tau}) can be written as
\begin{equation}g_{2}(d,\tau)=\langle\tilde{\psi}(0)|\hat{\Psi}_{+}^{\dagger}(d,\tau)\hat{\Psi}_{+}(d,\tau)|\tilde{\psi}(0)\rangle/|\langle\tilde{\psi}(0)|\tilde{\psi}(0)\rangle|^2.\end{equation}
The numerator describes the expectation value for the intensity
operator
$\hat{I}(\tau)=\hat{\Psi}_{+}^{\dagger}(d,\tau)\hat{\Psi}_{+}(d,\tau)$
in the Heisenberg picture given an initial state
$|\tilde{\psi}(0)\rangle$. However, we can easily convert this to
the Schr\"odinger picture by moving the evolution from the operator
to the state, \textit{i.e.}, by evolving $|\tilde{\psi}(0)\rangle$
under the same evolution equations
(Eqs.~\ref{eq:two_photon}-\ref{eq:one_photon}) and boundary
conditions (Eq.~\ref{eq:boundary_condition}) that we used earlier.
Therefore, the correlation function $g_{2}(z,\tau)$ will be given
by:

\begin{equation}
g_{2}(z,\tau)=\frac{\langle\tilde{\psi}(\tau)|\Psi_{+}^{\dagger}(z)\Psi_{+}(z)|\tilde{\psi}(\tau)\rangle}{\left|\langle\psi(0)|\Psi_{+}^{\dagger}(z)\Psi_{+}(z)|\psi(0)\rangle\right|^{2}}.\end{equation}

\section{Analytical solution for NLSE with open boundaries\label{sec:Analytical}}

In this section, we show that a NLSE system with open boundary
conditions yields analytical solutions in absence of an outside
driving source ($\alpha_0=0$). To obtain the analytical solutions, we use the Bethe ansatz technique~\cite{lieb-liniger,Lai:1989b}. This ansatz specifies that the eigenstates consist of a superposition of
states in which colliding particles exchange their wavenumbers
$k_i$. Unlike the typical formulation, the values of $k_{i}$ here
can be complex to reflect the \textit{open} nature of our boundary
conditions, which allow particles to freely enter or leave. In particular, we present the one-, two- and
many-body eigenmodes of the system along with their energy
spectra. Finding certain eigenmodes of the system (\textit{e.g.},
bound states) helps us understand the correlation functions and
also spatial wavefunctions which are numerically calculated later
in Sec.\ref{sec:Results} for a driven system.

\subsection{One-particle problem\label{sec:one_body}}

First, we calculate the fundamental modes for the one-particle
states. These modes are of particular interest when we later want
to construct the many-body wavefunction of the interacting system
in the absence of an input field.

Specifically, we want to find solutions of the Schr\"odinger
equation for a single particle in a system of length $d$,

\begin{equation}
i\frac{\partial}{\partial
t}\theta(z,t)=-\frac{1}{2m}\frac{\partial^{2}}{\partial
z^{2}}\theta(z,t),\end{equation}
subject to open boundary conditions. The boundary condition for
the undriven system at $z=0$ is given by

\begin{equation}
\theta(0)-\frac{i}{2m}\partial_{z}\theta(0)=0\end{equation}
 and similarly for $z=d$, \begin{equation}
\theta(d)+\frac{i}{2m}\partial_{z}\theta(d)=0.\end{equation} We
look for stationary solutions of the form $\theta(z,t)=e^{-i\delta
t}\theta(z)$, where $\theta(z)=A\sin(kz)+B\cos(kz)$. For
simplicity, we assume $m=1/2.$ Therefore, we recover the quadratic
dispersion relation $\delta=k^{2}$. The values of $k$ are allowed
to be complex to reflect the open nature of our boundary
conditions, which allows particles to freely enter or leave. By
enforcing the boundary conditions we get a set of equations for
the coefficients $A,B$,
\begin{eqnarray}
&~&B-iAk=0,\nonumber\\
&~&(A-iBk)\sin(kd)+(B+iAk)\cos(kd)=0,\nonumber\end{eqnarray}
which yields the characteristic equation for finding eigenmodes
and eigen-energies of system,

\begin{equation}
e^{2ikd}=\left(\frac{k+1}{k-1}\right)^{2}.\label{Eq:trans_noninteracting}\end{equation}
Therefore the normalized corresponding wave function for each
allowed \emph{k} will be:

\begin{eqnarray}
\theta(z)&=&A(\sin(kz)+ik\cos(kz)),\\
A^{2}&=&\frac{4k}{2dk(1+k^{2})+(k^{2}-1)\sin(2dk)}\nonumber\end{eqnarray}
We note that in the limit of large optical density $d\gg1$, the
lowest energy modes of the open system are very close to those of
a system with closed boundary conditions, whose characteristic
equation is given by $kd=n\pi$. For example, at $d=100$, the wave
number corresponding to lowest energy is
$k{\simeq}0.0314-i0.00063\simeq\pi/100$. We note that the
many-body solutions of the system in the presence of very strong
interactions~(large $\kappa$) can be constructed from these
single-particle solutions and proper symmetrization, as we show in
Sec.\ref{sec:many_body}.

\subsection{Two-particle problem \label{sec:two_body}}

In this section, we study the problem of two particles obeying the
NLSE with mixed boundary conditions. We wish to solve
\begin{eqnarray}
E\phi(z_{1},z_{2})&=&-\frac{1}{2m}\left(\frac{\partial^{2}}{\partial
z_{1}^{2}}+\frac{\partial^{2}}{\partial
z_{2}^{2}}\right)\phi(z_{1},z_{2})\nonumber\\
&+& 2\kappa\phi(z_{1},z_{2})\delta(z_{1}-z_{2}),\label{Eq:two_body}\end{eqnarray}
where $E$ is the energy of the system and can be complex. Again,
we assume the mass is entirely real, $m=1/2$.

We should note that the conventional method of separation of
variables cannot be applied in this case. The reason for this can
be understood in the following way. On one hand, if we ignore the
delta interaction term in the evolution equation of the two
particles, finding the eigenfunctions is essentially equivalent to
solving the Laplace equation in a box with mixed boundary
conditions. Therefore, for this problem the natural separation of
variables involves solutions given by products of functions
$f(z_{1})$ and $g(z_{2})$. On the other hand, if we neglect the
boundaries, the problem of two particles interacting at short
range can be solved by utilizing the center of mass and relative
coordinates and invoking solutions involving products of functions
$\tilde{f}(z_{1}+z_{2})$ and $\tilde{g}(z_{1}-z_{2})$. We
immediately see that the two sets of solutions are irreconcilable
and thus separation of variables is not applicable when both the
boundary conditions and interaction term are present.

We thus take a different approach, using a method similar to the
Bethe ansatz method for continuous, one-dimensional
systems~\cite{lieb-liniger}. Specifically, we solve the
Schr\"odinger equation in the triangular region where $0\leq
z_{1}<z_{2}\leq d$, and we treat the interaction as a boundary
condition at $z_{1}=z_{2}$. In other words, when two particles
collide with each other at $z_{1}=z_{2}$, they can exchange
momenta, which is manifested as a cusp in the wave function at
$z_{1}=z_{2}$. Hence, for the boundary conditions in this
triangular region, we have

\begin{eqnarray}
\phi(0,z_{2})-i\partial_{z_{1}}\phi(0,z_{2}) & = & 0\\
\phi(z_{1},d)+i\partial_{z_{2}}\phi(z_{1},d) & = & 0\\
\left(\partial_{z_{2}}-\partial_{z_{1}}\right)\phi(z_{1},z_{2})|_{z_{2}=z_{1}}
& = & \kappa\phi(z_{1},z_{2})|_{z_{2}=z_{1}}\end{eqnarray} We note
that the last boundary condition is deduced from integrating
Eq.~(\ref{Eq:two_body}) across $z_1=z_2$ and enforcing that the
wavefunction is symmetric,
\begin{eqnarray}
\left(\partial_{z_{2}}-\partial_{z_{1}}\right)\phi(z_{1},z_{2})|_{z_{2}=z_{1}^{+}}&-&\left(\partial_{z_{2}}-\partial_{z_{1}}\right)\phi(z_{1},z_{2})|_{z_{2}=z_{1}^{-}}\nonumber \\ &=&2\kappa\phi(z_{1},z_{2})|_{z_{1}=z_{2}}.\end{eqnarray}
Inside the triangle, the solution consists of superpositions of
\textit{free} particles with complex momenta. Since particles can
exchange momenta when they collide at $z_1=z_2$, we should
consider solutions of the following form,

\begin{equation}
\phi(z_{1},z_{2})=\sum_{\{\epsilon\}}\mathcal{A}_{\epsilon}e^{i\epsilon_{1}k_{1}z_{1}+i\epsilon_{2}k_{2}z_{2}}+\mathcal{B}_{\epsilon}e^{i\epsilon_{1}k_{2}z_{1}+i\epsilon_{2}k_{1}z_{2}}\end{equation}
where the summation should be performed on all sets of signs
$\epsilon=\pm1$. Given the terms containing
$\mathcal{A}_{\epsilon}$, the terms $\mathcal{B}_{\epsilon}$ then
arise from the scattering of the particles off each other. Let's
first consider the portion of the wavefunction containing the
terms $\mathcal{A}_{\epsilon}$, which we can write in the form:
\begin{eqnarray}
\phi_{\mathcal{A}}(z_{1},z_{2})&=&e^{ik_{1}z_{1}+ik_{2}z_{2}}+\alpha
e^{-ik_{1}z_{1}+ik_{2}z_{2}} \\
&+& \beta e^{-ik_{1}z_{1}-ik_{2}z_{2}}+\gamma
e^{+ik_{1}z_{1}-ik_{2}z_{2}},\nonumber \end{eqnarray} where the energy is
equal to $E=k_{1}^{2}+k_{2}^{2}$ and could be complex. Similar to
the single-particle solutions, the presence of the imaginary part
in the energy reflects the fact that the two-particle state stays
a finite amount of time inside the system. Applying boundary
conditions at $z=0$ and $z=d$ subsequently generates four
equations relating $\alpha,\beta,\gamma$ where one of them is
redundant. Their solution reduces the wavefunction to
\begin{eqnarray}
\phi_{\mathcal{A}}(z_{1},z_{2})&=&e^{ik_{1}z_{1}+ik_{2}z_{2}}+\frac{k_{1}+1}{k_{1}-1}e^{-ik_{1}z_{1}+ik_{2}z_{2}}\\
&+&\frac{k_{1}+1}{k_{1}-1}\gamma
e^{-ik_{1}z_{1}-ik_{2}z_{2}}+\gamma e^{+ik_{1}z_{1}-ik_{2}z_{2}}\nonumber \end{eqnarray} where
$\gamma=\frac{k_{2}-1}{k_{2}+1}e^{2ik_{2}d}$. A similar expression
results for the portion of $\phi(z_1,z_2)$ containing the
$\mathcal{B}_{\epsilon}$ terms, once the boundary conditions at
$z=0$ and $z=d$ are applied:
\begin{eqnarray*}
\phi_{\mathcal{B}}(z_{1},z_{2})&=&\frac{1}{t}\left(e^{ik_{2}z_{1}+ik_{1}z_{2}}+\frac{k_{2}+1}{k_{2}-1}e^{-ik_{2}z_{1}+ik_{1}z_{2}}\right)\\
&+&\frac{\gamma'}{t}\left(\frac{k_{2}+1}{k_{2}-1}
e^{-ik_{2}z_{1}-ik_{1}z_{2}}+e^{+ik_{2}z_{1}-ik_{1}z_{2}}\right)\end{eqnarray*}
where $\gamma'=\frac{k_{1}-1}{k_{1}+1}e^{2ik_{1}d}$, and $t$ is a
coefficient to be determined from the boundary condition at
$z_1=z_2$. To find $t$, it is convenient to re-write each of ther
terms in $\phi_{\mathcal{A,B}}$ as a product of relative
coordinate~($r=z_{2}-z_{1}$) and center-of-mass
coordinate~($R=(z_{1}+z_{2})/2$) functions,
\begin{eqnarray}
\tilde{\phi}_{\mathcal{A}}(R,r)&=&e^{ipR-iqr}+\frac{k_{1}+1}{k_{1}-1}e^{-iqR+ipr}\\
&+&\gamma\frac{k_{1}+1}{k_{1}-1}e^{-ipR+iqr}+\gamma
e^{iqR-ipr},\end{eqnarray}
\begin{eqnarray}
\tilde{\phi}_{\mathcal{B}}(R,r)&=&\frac{1}{t}(e^{ipR+iqr}+\frac{k_{2}+1}{k_{2}-1}e^{iqR+ipr}\\
&+&\gamma'\frac{k_{2}+1}{k_{2}-1}e^{-ipR-iqr}+\gamma'e^{-iqR-ipr}),\end{eqnarray}
where $p=(k_{1}+k_{2})$ and $q=(k_{1}-k_{2})/2$.
The boundary condition at $z_1=z_2$ leaves the center-of-mass
parts of the wavefunction unaffected, but yields the following
condition on the relative coordinates,
\begin{equation}
\partial_{r}\phi(R,r)|_{r=0^{+}}=\frac{\kappa}{\sqrt{2}}\phi(R,r)|_{r=0^{+}}.\end{equation}
where $\phi=\phi_{\mathcal{A}}+\phi_{\mathcal{B}}$ is the total
wavefunction in the triangular region. We should satisfy this
boundary condition separately for each of the center-of-mass
momentum terms $e^{{\pm}ipR},e^{{\pm}iqR}$ in the total
wavefunction. This leads to three independent equations (one out
of four is redundant). However, we introduce a new parameter
($t'$) to simplify the equations, which turns them into four
equations:

\begin{eqnarray}
t=\frac{k_{1}-k_{2}+i\kappa}{k_{1}-k_{2}-i\kappa}~~ & , & ~~t'=\frac{k_{1}+k_{2}+i\kappa}{k_{1}+k_{2}-i\kappa}\label{Eq:trans1}\\
t~t'\left(\frac{k_{1}+1}{k_{1}-1}\right)^{2} & = & e^{2ik_{1}d}\label{Eq:trans2}\\
t'\left(\frac{k_{2}+1}{k_{2}-1}\right)^{2} & = &
te^{2ik_{2}d}\label{Eq:trans3},\end{eqnarray} which can be written
in the following short form:

\begin{equation}
e^{2ik_{i}d}=\frac{(k_{i}+1)^{2}}{(k_{i}-1)^{2}}\prod_{j\neq
i}\frac{(k_{i}-k_{j}+i\kappa)(k_{i}+k_{j}+i\kappa)}{(k_{i}-k_{j}-i\kappa)(k_{i}+k_{j}-i\kappa)}\label{eq:transcedent}\end{equation}
where $i,j$ can be (1,2). These are transcendental equations for
($k_{1},k_{2}$), which generate the spectrum of two interacting
particles. We can also write the wave functions
($\phi=\phi_{\mathcal{A}}(z_{1},z_{2})+\phi_{\mathcal{B}}(z_{1},z_{2})$)
in the region ($0\leq z_{1}<z_{2}\leq d)$ in a more compact way,
by using the single particle solutions
$\eta_{k}(z)=\sin(kz)+ik\cos(kz)$:

\begin{eqnarray}
\phi_{\mathcal{A}}(z_{1},z_{2}) & = & \frac{4}{k_{1}-1}\frac{e^{ik_{2}d}}{k_{2}+1}\eta_{k_{1}}(z_{1})\eta_{k_{2}}(d-z_{2})\\
\phi_{\mathcal{B}}(z_{1},z_{2}) & = &
\frac{4t^{-1}}{k_{2}-1}\frac{e^{ik_{1}d}}{k_{1}+1}\eta_{k_{2}}(z_{1})\eta_{k_{1}}(d-z_{2}).\end{eqnarray}

It is interesting to note that in the limit of strong
interaction (either for positive or negative $\kappa$), the
solutions are very similar to the non-interacting case. The reason
can be seen from the transcendental Eqs. (\ref{eq:transcedent}),
in the limit $\kappa\rightarrow\pm\infty$. We then recover the
same characteristic equations $e^{2ikd}=(\frac{k+1}{k-1})^{2}$ for
both wavevectors as the non-interacting case,
Eq.~(\ref{Eq:trans_noninteracting}). We should note that there are some trivial solutions to the
transcendental Eqs. (\ref{eq:transcedent}), which do not have any
physical significance. For example, equal wave vectors
$k_{1}=k_{2}$. Although one can find such wave vectors, this
solution is readily not a solution to Eq.(\ref{Eq:two_body}),
since it does not satisfy the interacting part (this solution only
contains center of mass motion). One can also plug back the wave
vectors into wave function and arrive at a wave function equal to
zero everywhere. Another example is when one of the wave vectors
is zero. In this case, one can also show that the wave function
is zero everywhere. If $(k_{1},k_{2})$ are solutions to the
transcendental equations, then $(\pm k_{1},\pm k_{2})$ are also
solutions with equal energies. In next two sections, we investigate non-trivial solutions to the transcendental equation for two particles and discuss the related physics. 

\subsection{Solutions close to non-interacting case} 

\begin{figure}[h]
\centering
\includegraphics[width=0.45\textwidth]{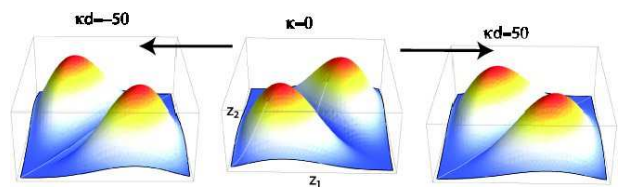}

\caption[ Two-photon wave function with different modes]{The amplitude of the two-photon wavefunction for (m=1, n=2) mode when
the system is not driven. By increasing the interaction, photons self-organize
inside the medium and exhibit anti-bunching (depletion of diagonal
elements). For this plot: $d=30$.}

\label{fig:wavefunction_different_modes}
\end{figure}

The transcendental equations allow a set of solutions
with the wavevectors close to two different non-interacting modes
say $(m,n)$. In the non-interacting regime, any mode can be
populated by an arbitrary number of photons. However, once the
interaction is present, photons can not occupy the same mode and
therefore, the photons will \textit{reorganize} themselves and each acquire
different modes.  Fig. \ref{fig:wavefunction_different_modes} shows
a normal mode wavefunction of a non-driven system in both
non-interacting and strongly interacting regime ($\kappa d\gg1$).
The wave function has a cusp on its diagonal and diagonal elements are depleted for both repulsive
and attractive strong interaction.

\begin{figure}[h]
\includegraphics[width=.4\textwidth]{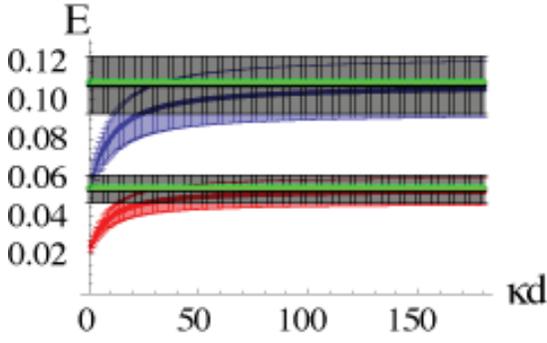}
\caption[Energy of two-photon states occupying different modes]{Energy of two-photon states: by increasing the interaction strength
($\kappa d\gg1$) the energy of interacting particles (red:E(1,2)
and blue:E(1,3)) reaches that of the non-interacting particles (black).
For large system (in this case $d=30\gg1$), the energy limit is equal
to energy of particles in a closed box (green). The error bars show
that imaginary part of the energies.}

\label{fig:energy_different_modes}
\end{figure}

This is a manifestation of \textit{fermionization} of bosons in one dimensional system in the presence of strong interaction \cite{Girardeau:1960,lieb-liniger}.
Such solutions can exist both for repulsive and attractive interactions. However, we note in the case of
attractive interaction such solutions are not the ground state of
the system and solutions with lower energies exist which will be
discussed below. We later argue that indeed on the repulsive side, the anti-bunching
behavior of a driven system is due to the repulsion of the photons
inside the medium. We can also estimate the energy of such modes which is always positive.
In the strong interacting regime, particles avoid each other and
therefore, their energy of a strongly two interacting bosons
$E(m,n)$ will be equal to the energy of a system which has two
non-interacting bosons, one in state $m$ and the other in state
$n$. This is shown in Fig. \ref{fig:energy_different_modes}, where
by increasing the interaction strength the energy of interacting
particles reaches that of the non-interacting particles. As we
pointed out in the previous section (Sec.\ref{sec:one_body}), the
energy of modes ($E(n)$) in an open box has an imaginary part
which represents how fast the particle leave the system. However,
for large systems ($d\gg1$), this decay is very small compared to
the energy of the mode and one can approximate the energy of an
open system by that of a closed box (i.e.
$E(n)\simeq(\frac{n\pi}{d})^{2}$). Therefore, the energy of two
strongly interacting photons ($\kappa d\gg1$), in the limit of
large system ($d\gg1$), will be given by: \begin{equation}
E(n,m)\simeq\left(\frac{n\pi}{d}\right)^{2}+\left(\frac{m\pi}{d}\right)^{2}\end{equation}

 We note that our strongly interacting system is characterized by the parameter $\kappa d$ which
is the same $\gamma$-parameter conventionally used for interacting
1D Bose gas. More precisely, the $\gamma$-parameter which is the
ratio of the interaction to kinetic energy can be simplified in
our case for two particles: $m\kappa d/2=\kappa d/4$.

\subsection{Bound States Solution} 

\begin{figure}[h]
\centering
\includegraphics[width=0.4\textwidth]{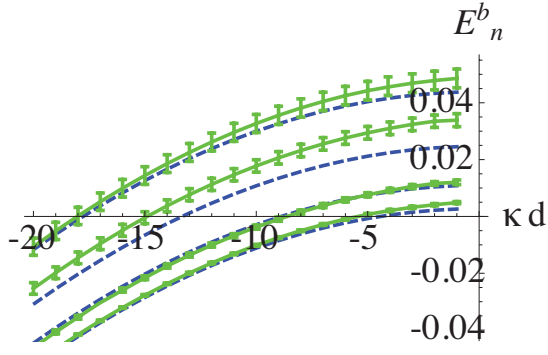}

\caption[Energy of bound states versus strength of nonlinearity(2)]{Energy of bound states versus strength of nonlinearity. Green (solid)
curves are obtained by solving transcendental Eqs. (\ref{eq:transcedent}).
Blue (dashed) curve are estimated based on $E_{n}^{b}\simeq\left(\frac{n\pi}{d}\right)^{2}-\frac{\kappa^{2}}{2}$.
In this plot $d=30$. }

\label{fig:energy_bound_state_single}
\end{figure}

For attractive interaction ($\kappa<0$), the mode
equation~(\ref{eq:transcedent}) admits solutions which take the
form of photonic bound states. Specifically, in the reference
frame of the center of mass, two particles experience an
attractive delta function interaction
$-2|\kappa|\delta(z_{2}-z_{1})\rightarrow-\sqrt{2}|\kappa|\delta(r)$,
which allows one bound state in the relative coordinate.
Therefore, the part of the wavefunction describing the relative
coordinate roughly takes the form $e^{iq|r|}$, where the relative
momentum $q=(k_{1}-k_{2})/\sqrt{2}\simeq i|\kappa|/\sqrt{2}$ is
imaginary and its energy is about $-\kappa^{2}/2$. On the other
hand, the center of mass momentum can take a discrete set of
values that are determined by the system boundary conditions. We
find that the center of mass solutions can be approximately
described by two different types. The first type is where the real
part of each photon wavevector roughly takes values allowed for a
single particle in a box, such that
$k_{1}\simeq\left(\frac{n\pi}{d}\right)+i\frac{\kappa}{2}$ and
$k_{2}\simeq\left(\frac{n\pi}{d}\right)-i\frac{\kappa}{2}$. In
this case the center of mass has wavevector
$p=k_{1}+k_{2}\simeq2(\frac{n\pi}{d})$. The
corresponding energies for these states are

\begin{equation}
E_{n}^{b}\simeq2\left(\frac{n\pi}{d}\right)^{2}-\frac{\kappa^{2}}{2}.\label{eq:bound_energy_double}\end{equation}
Here, the first term on the right corresponds to the energy of the
center of mass motion, and the second term corresponds to the
bound-state energy of the relative motion. Fig.
\ref{fig:energy_bound_state_single} shows that the energies
estimated in this way agree very well with the exact values
obtained by solving the transcendental Eqs.
(\ref{eq:transcedent}).

\begin{figure}[h]
\centering
\includegraphics[width=0.40\textwidth]{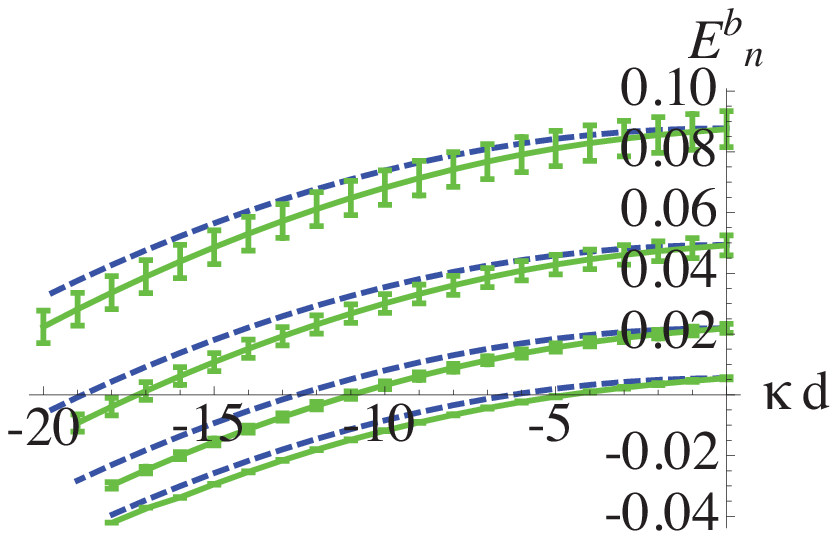}

\caption[Energy of bound states versus strength of nonlinearity (1)]{Energy of bound states versus strength of nonlinearity. Green (solid)
curves are obtained by solving transcendental Eqs. (\ref{eq:transcedent}).
Blue (dashed) curves are estimated based on $E_{n}^{b}\simeq2\left(\frac{n\pi}{d}\right)^{2}-\frac{\kappa^{2}}{2}$.
In this plot $d=30$. }

\label{fig:energy_bound_state_double}
\end{figure}
\begin{figure}[h]
\centering
 \includegraphics[width=0.45\textwidth]{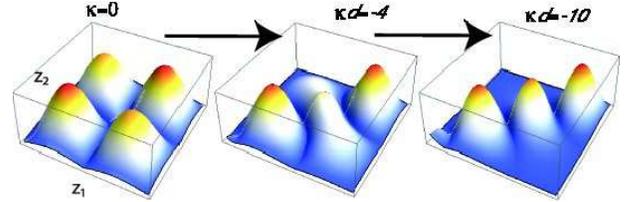}

\caption[ Two-photon wave function of a bound state (1)]{The amplitude of the two-photon wavefunction for a bound state when
the system is not driven. By increasing interaction, photons becomes
more bunched. $(k_{1},k_{2})\simeq\left(\frac{n\pi}{d}\right)\pm i\frac{\kappa}{2}$.
For this plot: $d=60,n=3$.}

\label{fig:wavefunction_bound_state_double}
\end{figure}

The second type of solution allowed for the center of mass motion
is where its energy approximately takes a single-particle
value, $\frac{p^2}{2(2m)}=(\frac{n\pi}{d})^2$ where $p=k_{1}+k_{2}\simeq\sqrt{2}(\frac{n\pi}{d})$.
Therefore, the momentum of individual particles will be given by
$k_{1}\simeq\left(\frac{n\pi}{\sqrt{2}d}\right)+i\frac{\kappa}{2}$,
$k_{2}\simeq\left(\frac{n\pi}{\sqrt{2}d}\right)-i\frac{\kappa}{2}$
and the energy of this paired composite can be estimated as
\begin{equation}
E_{n}^{b}\simeq\left(\frac{n\pi}{d}\right)^{2}-\frac{\kappa^{2}}{2}.\label{eq:bound_energy_single}\end{equation}
Again, the estimated energies agree well with exact solutions, as
shown in Fig. \ref{fig:energy_bound_state_single}. We note that
some of the estimated allowed energies for the two types of center
of mass solutions coincide (\textit{e.g.}, the lowest lying energy
level in Fig. \ref{fig:energy_bound_state_double} and
Fig. \ref{fig:energy_bound_state_single}). 

The energies of this series of bound states decrease with
increasing strength of nonlinearity $|\kappa|$. Now, suppose we
drive the system with a coherent field of fixed frequency
$\delta$, while varying $\kappa$.  The system is expected to
display a set of resonances as $|\kappa|$ is increased, each time
$\delta$ is equal to some particular bound state energy $E_n^b$.
This effect in fact gives rise to oscillatory behavior in the
correlation functions as a function of $\kappa$, as we will see
later (Fig. \ref{fig:attractive}(a) and (b)).

\begin{figure}[h]
\centering
\includegraphics[width=0.45\textwidth]{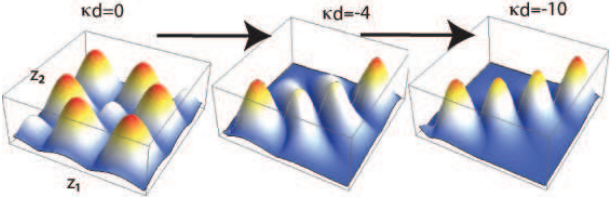}

\caption[ Two-photon wave function of a bound state (2)]{The amplitude of the two-photon wavefunction for a bound state for
a non-driven system. By increasing interaction, photons becomes more
bunched. $(k_{1},k_{2})\simeq\left(\frac{n\pi}{\sqrt{2}d}\right)\pm i\frac{\kappa}{2}$.
For this plot: $d=60,n=3$.}

\label{fig:wavefunction_bound_state_single}
\end{figure}

The wavefunction amplitude of a typical bound state is shown in
Fig. \ref{fig:wavefunction_bound_state_single}. Due to the
attractive interaction, diagonal elements $z_1=z_2$ become more
prominent as $|\kappa|$ increases, indicating a stronger bunching
effect for the photons, and these states become more tightly bound
in the relative coordinate. The center of mass of the bound states
can acquire a free momentum that is quantized due to the system
boundary conditions~(\textit{e.g.}, $k\simeq n\pi/d$).
Fig. \ref{fig:wavefunction_bound_state_single} shows the
wavefunction of the third bound state (n=3). The three peaks
evident for large $|\kappa|$ reflect the quantum number of the
center of mass motion.

\subsection{Many-body problem\label{sec:many_body}}

In this section, we obtain the general solution for the many-body
case. For the many-body system, the Schr\"odinger equation takes the
form

\begin{eqnarray}
E\phi(z_{1},...,z_{N})&=&-\frac{1}{2m}\sum_{i}\frac{\partial^{2}}{\partial
z_{i}^{2}}\phi(z_{1},...,z_{N})\\
&+&\sum_{<i,j>}2\kappa\phi(z_{1},...,z_{N})\delta(z_{i}-z_{j}),\label{Eq:many_body}\end{eqnarray}
where $<i,j>$ indicates pairs of particles. The open boundary
conditions for the many-body problem are given by

\begin{eqnarray}
\left[\phi(z_{1},...,z_{N})-i\frac{\partial}{\partial z_{i}}\phi(z_{1},...,z_{N})\right]_{z_{i}=0} & = & 0\label{eq:boundary_many_a}\\
\left[\phi(z_{1},...,z_{N})+i\frac{\partial}{\partial
z_{i}}\phi(z_{1},..,z_{N})\right]_{z_{i}=d} & = &
0.\label{eq:boundary_many_b}\end{eqnarray} Before presenting the
general many-body solution, we first study the limit of very large
interaction strength for two particles. In the limit of hardcore
bosons where $\kappa\rightarrow\infty$, the expressions can be
simplified since $t,t'=-1$ and $e^{ikd}=\frac{k+1}{k-1}$ for both
$k=k_{1,2}$. Then, the two components of the wavefunction
$\phi_{\mathcal{A}}$ and $\phi_{\mathcal{B}}$ take very similar
forms,

\begin{eqnarray}
\phi_{\mathcal{A}}(z_{1},z_{2}) & = & \frac{4}{(k_{1}-1)(k_{2}-1)}\eta_{k_{1}}(z_{1})\eta_{k_{2}}(z_{2})\\
\phi_{\mathcal{B}}(z_{1},z_{2}) & = &
\frac{-4}{(k_{1}-1)(k_{2}-1)}\eta_{k_{2}}(z_{1})\eta_{k_{1}}(z_{2}).\end{eqnarray}
The generalization to the many-body solution is straightforward
for the hardcore boson case (also see Ref.~\cite{Kojima:1997}):
\begin{equation}
\phi(z_{1},z_{2},...,z_{N})=\left(\prod_{j=1}^{N}\frac{1}{(k_{j}-1)}\right)\left|\det_{1\leq
j,k\leq N}\eta_{j}(z_{k})\right|\end{equation} Similar to two-body
solution, we note that such solutions are present both for
positive and negative $\kappa$. Since the system is one
dimensional, strong interaction leads to fermionization of bosons
( in this case photons)\cite{Girardeau:1960,lieb-liniger}.

We can also extend the many solution for an arbitrary interaction
strength, following Refs.\cite{lieb-liniger, Bulatov:1988}. Similar to the two-body
case, we can construct the general many-body wave function of the
form: \begin{equation}
\phi(z_{1},z_{2},...,z_{N})=\sum_{\epsilon}A_{\epsilon}\sum_{P}B_{P}e^{i\sum_{i}\epsilon_{p_{i}}k_{p_{i}}z_{i}}\end{equation}
where the first sum is over forward and backward going waves $(\epsilon=\pm$1)
and the second sum is over different momentum permutations of the
set $\left\{ k\right\} =(k_{1},k_{2},...,k_{N})$, therefore there
are $2^{N}N!$ terms. We can find $B_{P}$ coefficient by requiring
$\sum_{P}B_{P}\prod_{i<j}e^{i\epsilon_{p_{i}}k_{p_{i}}z_{i}}$ to
be solution to the Schr\"odinger equation (Eq.\ref{Eq:many_body}).
We can write these coefficients in a compact way according to Gaudin \cite{Gaudin:1983,Dorlas:1993},
$B_{P}=\prod_{i<j}\left(1+\frac{i\kappa}{\epsilon_{p_{i}}k_{p_{i}}-\epsilon_{p_{j}}k_{p_{j}}}\right)$
with the total energy $E=\sum_{i}k_{i}^{2}$. Now, we apply the boundary
condition Eq.(\ref{eq:boundary_many_a}) which relates coefficient
$A_{\epsilon}$. For a given momentum permutation $P=(p_{1},p_{2,},...,p_{N})$,
by considering the terms corresponding to different signs of $\epsilon_{p_{i}}$,
the boundary condition requires $A_{\epsilon}$ to satisfy equations
of the form\begin{eqnarray*}
(1+\epsilon_{p_{i}}k_{p_{i}})A_{\epsilon_{1},..\epsilon_{p_{i}}..\epsilon_{N}}\prod_{j(\neq p_{i})}\left(1+\frac{i\kappa}{\epsilon_{p_{i}}k_{p_{i}}-\epsilon_{j}k_{j}}\right) & +\\
(1-\epsilon_{p_{i}}k_{p_{i}})A_{\epsilon_{1},..(-\epsilon_{p_{i}})..\epsilon_{N}}\prod_{j(\neq p_{i})}\left(1+\frac{i\kappa}{-\epsilon_{p_{i}}k_{p_{i}}-\epsilon_{j}k_{j}}\right) & = & 0\end{eqnarray*}
The above equations can be satisfied by the following solution for
$A_{\epsilon}$ \begin{equation}
A_{\epsilon}=\prod_{i<j}\left(1-\frac{i\kappa}{\epsilon_{i}k_{i}+\epsilon_{j}k_{j}}\right)\prod_{m=1}^{N}\left(1-\frac{1}{\epsilon_{m}k_{m}}\right).\end{equation}
Therefore, the wavefunction can be written as: \begin{eqnarray}
&\phi&(z_{1},z_{2},...,z_{N})=\\
&~&\sum_{\epsilon}\tiny{\sum_{P}}\prod_{m=1}^{N}\left(1-\frac{1}{\epsilon_{m}k_{m}}\right)e^{[i(\epsilon_{p_{1}}k_{p_{1}}x_{1}+..+\epsilon_{p_{N}}k_{p_{N}}x_{N})]}\nonumber \\
&\times& \prod_{i<j}\left[\left(1-\frac{i\kappa}{\epsilon_{i}k_{i}+\epsilon_{j}k_{j}}\right)\left(1+\frac{i\kappa}{\epsilon_{p_{i}}k_{p_{i}}-\epsilon_{p_{j}}k_{p_{j}}}\right)\right]\nonumber\end{eqnarray}
 Similar to two-body case, we have to subject this solution to the
boundary condition at other end (i.e., $z=d$) to determine the momenta
$k_{i}$'s. This condition yields the transcendental equations for
momenta: \begin{equation}
e^{2ik_{i}d}=\frac{(k_{i}+1)^{2}}{(k_{i}-1)^{2}}\prod_{j\neq i}\frac{(k_{i}-k_{j}+i\kappa)(k_{i}+k_{j}+i\kappa)}{(k_{i}-k_{j}-i\kappa)(k_{i}+k_{j}-i\kappa)}.\end{equation}
 If we assume only two particles in the system, one can easily verify
the the above transcendental equations reduce to two-body transcendental
equation derived in the previous section (Eq.(\ref{eq:transcedent})).

\section{Quantum transport properties \label{sec:Results}}

In this section, we investigate transport properties of the
photonic nonlinear one-dimensional system in the regimes of
attractive, repulsive, and absorptive interactions between
photons. We present numerical solutions for the transport of
photons incident from one end of the waveguide~(a driven system),
while using the analytical solutions of the non-driven system
(Sec.~\ref{sec:Analytical}) to elucidate the various behaviors that
emerge in the different regimes.

\subsection{Repulsive Interaction $(\kappa>0)$}

We first study the quantum transport properties of the system in
the dispersive regime where the nonlinearity coefficient is almost
real and positive $(\kappa>0)$, such that photons effectively
\textit{repel} each other inside the system.

\begin{figure}[h]
\centering
\includegraphics[width=0.4\textwidth]{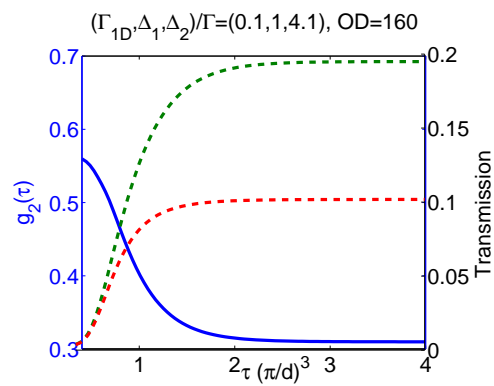}

\caption[Reaching the steady-state: dispersive case]{$g_{2}(\tau=0)$ reaches the steady-state after a time interval which
is set by the bandwidth of the system, one-photon state (green) is
partially transmitted while the transmission of the two-photon state
(red) is further suppressed due the nonlinear dispersion. This has
been generated for a system with $\Gamma_{1D}/\Gamma=10\%$ an OD=160.\label{fig:filter_dispersive}}

\end{figure}

We assume that a weak coherent field is incident to the waveguide
at one end, $z=0$, with no input at the other end,
$z=d$~{[}similar to Fig. \ref{fig:four_level}(b)]. We fix the
detuning of the input field to $\delta_{0}=(\pi/d)^{2}$, which
corresponds to the first transmission resonance in the linear
regime (Sec.~\ref{sec:linear}). Because we have assumed a weak
input field, we can apply the techniques described in
Sec.~\ref{sec:quantum_formalism} to describe the transport. Our
numerical techniques for solving these equations are given in
Appendix~\ref{sec:numerics}. While the numerical results presented
in this and the following sections are evaluated for a specific
set of parameters (system size, detuning, etc.), the conclusions
are quite general. Numerically, we begin with no photons inside
the medium, and evaluate quantities such as the transmission
intensity and correlation functions only after the system reaches
steady state in presence of the driving field. In Fig.\ \ref{fig:filter_dispersive}, the transmission of the single-photon intensity 
\begin{equation}
T_1=\frac{\langle1|\Psi^\dagger_+(d)\Psi_+(d)|1\rangle}{\langle1|\Psi^\dagger_+(0)\Psi_+(0)|1\rangle},\label{eq:T1}\end{equation} the transmission of the two-photon intensity  \begin{equation}T_2=\frac{\langle2|\Psi^\dagger_+(d)\Psi_+(d)|2\rangle}{\langle2|\Psi^\dagger_+(0)\Psi_+(0)|2\rangle}\label{eq:T2}\end{equation} and the transmitted correlation function $g_2(z=d,\tau=0)$ is shown as the system evolves in time. The system reaches its steady state after a time of the order of the inverse of the system bandwidth (Sec.~\ref{sec:linear}). In fact, $T_1$ coincides with the linear transmission coefficient of the system in the absence of the nonlinearity.

\begin{figure}[h]
\centering
\includegraphics[width=0.45\textwidth]{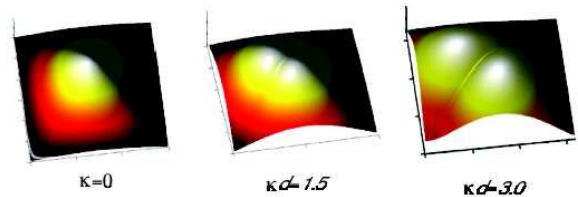}

\caption[Two-photon wave function: Repulsive case]{Two-photon wave function $|\phi(z_{1},z_{2})|$ exhibiting delocalization.
We have assumed no dissipation ($\Gamma'=\Gamma=0$) in this plot.
$d=30$ for different values of $\kappa$.\label{fig:quantum_repulsive}}
\end{figure}

First, we note that the single-photon wave function is not
affected by the presence of the nonlinearity and will be perfectly
transmitted in the absence of linear losses. Thus, in our
truncated Hilbert space, the only subspace affected by $\kappa$ is
the two-photon wave function, which is shown in
Fig. \ref{fig:quantum_repulsive}. We clearly observe that the nonlinearity causes
repulsion between two photons inside the system, as the wave
function along the diagonal $z_{1}=z_{2}$ becomes suppressed while
the off-diagonal amplitudes become peaked~(indicating the
de-localization of the photons). This behavior closely resembles
that of the natural modes of the system, as calculated in
Sec.~\ref{sec:Analytical}. A similar behavior involving the
``self-organization'' of photons in an NLSE system in
equilibrium has been discussed in Ref. \cite{Chang:2008}.

\begin{figure}[h]
\centering
\includegraphics[width=0.45\textwidth]{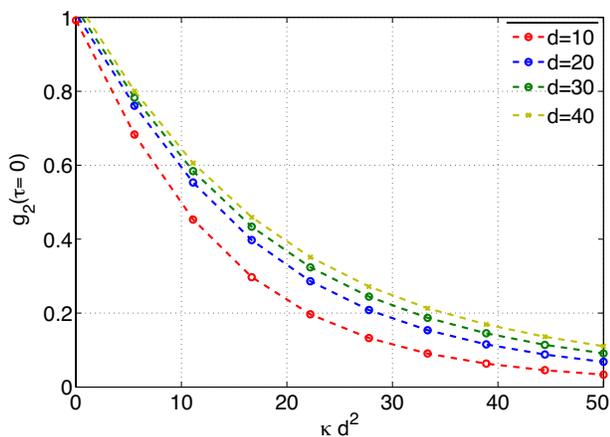}

\caption[ Scaling of $g_2$ with nonlinearity strength: repulsive case]{$g_{2}(\tau=0)$ as a function of nonlinearity. For large system sizes
$d\gg1$, the anti-bunching of the system scales with $\kappa d^{2}$.}
\label{fig:repulsive_scaling_d}
\end{figure}

In the presence of linear absorption (discussed in
Sec.~\ref{sec:linear}), the system will not be perfectly
transmitting even on resonance, and therefore in a realistic
situation the transmittivity will be less than one $(T_{1}<1)$.
Note, however, that such absorption would result in a classical
output given a classical input. Significantly, in the presence of
a nonlinearity, we find that the output light can acquire
non-classical character. Specifically, the transmitted light
exhibits anti-bunching~($g_{2}(z=d,\tau=0)<1$), which becomes more
pronounced with increasing ${\kappa}d^{2}$
(Fig. \ref{fig:repulsive_scaling_d}). This effect partly arises
from the suppression of transmission of two-photon components, due
to an extra nonlinear phase shift that shifts these components out
of transmission resonance. In fact, these components are more
likely to get reflected, which causes the reflected field to
subsequently exhibit bunching behavior. We note that this effect
is similar to photon blockade in a cavity (e.g., see
Refs.\cite{Imamoglu:1997,Grangier:1998,Imamoglu:1998}). In
addition, additional anti-bunching occurs due to the fact that
two-photon components inside the system tend to get repelled from
each other. This effect arises due to the spatial degrees of
freedom present in the system, which is fundamentally different
than switching schemes proposed in optical cavities (e.g.,
Refs.\cite{Imamoglu:1997,Grangier:1998,Imamoglu:1998}) or
wave-guides coupled to a point-like emitter
\cite{ShanhuiFan:PRL2007,Chang:2007}. In the limit where
$\kappa\rightarrow\infty$, the transmitted field approaches
perfect anti-bunching, $g_{2}(d,\tau=0)=0$.

\begin{figure}[h]
\centering
\includegraphics[width=.45\textwidth]{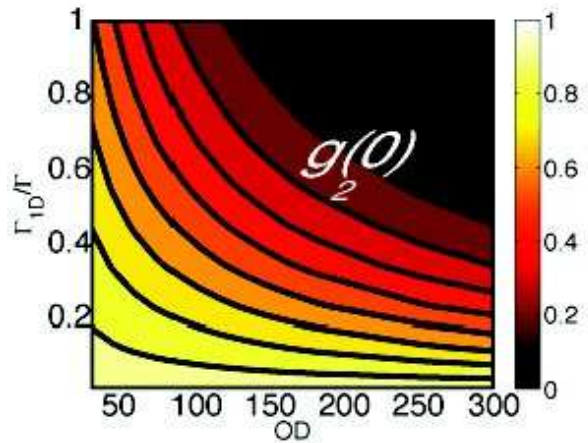}
\caption[Scaling of the anti-bunching with optical density and cooperativity]{Repulsive photons: Correlation
function $g_2(\tau=0)$ of the transmitted light when the frequency is set to the single-photon transmission resonance with
$T_1\simeq 20\%$ and $\frac{\Delta_2}{\Gamma}=5$.}\label{fig:repulsive_scaling_OD}
\end{figure}

In an experimental realization, the requirement to see the
photon repulsion ($\kappa d^{2}\geq40$) for a system with
$\Gamma_{1D}/\Gamma=10\%$, would be a coherent optical length of
$d\simeq40$ when $\Delta_{2}/\Gamma=1$. Therefore, at least an
optical density of $OD\simeq160$ is needed for $T_{1}\simeq20\%$. The anti-bunching in the transmitted light  is more pronounced as the optical density increases,
which increases the effective system finesse (Fig.~\
\ref{fig:repulsive_scaling_OD}).

\subsection{Attractive Interaction $(\kappa<0)$}

In this section, we study the quantum transport properties of the
system in the presence of dispersive nonlinearity with negative
coefficient. Contrary to the semi-classical prediction, we show
that the second-order correlation function of the transmitted
field oscillates as function of nonlinear interaction strength and
can exhibit both bunching and anti-bunching. We explain the origin
of this behavior in terms of the analytical solutions obtained in
Sec.~\ref{sec:two_body}.

\begin{figure}[h]
\centering
\includegraphics[width=0.45\textwidth]{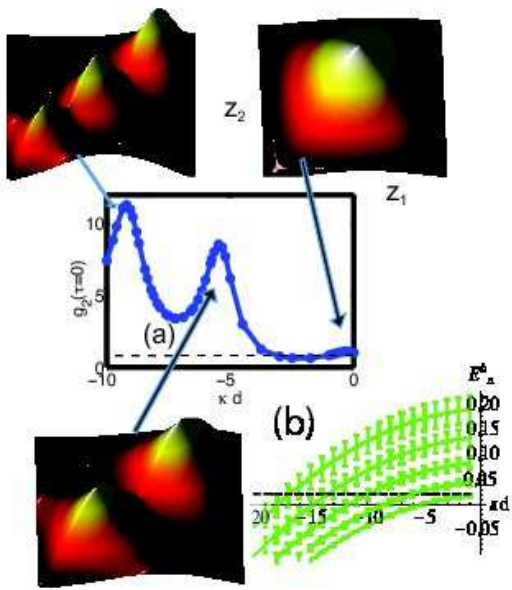}

\caption[Resonances in presence of attractive interaction]{(a) Output correlation function $g_{2}(\tau=0)$ as a function of
nonlinearity: When the negative nonlinear strength is changed to higher
values and $g_{2}$ exhibits resonances at certain values of $\kappa d\simeq(0,6,10,14,...)$.
In this plot the system size is $d=30$, however, for other system
size same behaviors were observed around similar values of $\kappa d$.
The two-photon wavefunction ($|\phi(z_{1},z_{2})|$) for four values
of nonlinearity is shown. (b) Corresponding bound state energies (green-solid)
which become resonant with incoming photon energy (black-dotted) for
specific nonlinearities. We have assumed no dissipation ($\Gamma'=\Gamma=0$)
in these plots. }

\label{fig:attractive}
\end{figure}

In Fig.\ \ref{fig:attractive}(a), we plot $g_{2}(\tau=0)$ for the
transmitted field versus ${\kappa}d$. Initially, the system
exhibits anti-bunching behavior for small values of $|\kappa|d$\,
which indicates that multi-photon components tend to switch
themselves out of transmission resonance. However, as we increase
$|\kappa|d$, oscillations develop in the correlation function,
exhibiting strong bunching behavior at particular values of
${\kappa}d$. Thus, unlike the repulsive case, a competing behavior
arises between the photon switching effect and the resonant
excitation of specific bound states within the system, as we
describe below. In particular, the bound state energies
$E_{n}^{b}$ decrease quadratically with changing $\kappa$,
according to Eq.\ (\ref{eq:bound_energy_single}) or Eq.\
(\ref{eq:bound_energy_double}), which is shown in Fig.
\ref{fig:attractive}(b). For a fixed detuning $\delta$, the
oscillation peaks~(where $g_{2}$ is largest) correspond to
situations where the energy of a bound state becomes equal to the
energy of two incoming photons ($E_{n}^{b}=2\delta$). This effect
is further confirmed by examining the two-photon wave function at
each of these oscillation peaks~(Fig.\ \ref{fig:attractive}a). We
clearly observe that these wave functions correspond to the bound
states calculated in Sec.~\ref{sec:two_body}. Similar to Fig.
\ref{fig:wavefunction_bound_state_double} and Fig.
\ref{fig:wavefunction_bound_state_single}, it is readily seen that
the wave functions at these peaks are localized along the
diagonal, indicating a bound state in the relative coordinates and
leading to the bunching effect in transmission. On the other hand,
an increasing number of nodes and anti-nodes develop along the
diagonal for increasing $|\kappa|d$, which are associated with the
higher momenta of the center-of-mass motion. We note that such
resonances deviate significantly from the the semiclassical
picture, where anti-bunching was predicted for both positive and
negative nonlinearity. We also note that in cavity QED systems
this effect is not present since these systems are single-mode.

The experimental requirement to see such behaviors is more
stringent than the photon repulsion in the previous section. For
example, if we want to observe the second photonic bound state
($\kappa d\geq5$) for a system with $\Gamma_{1D}/\Gamma=0.2$, the
coherent optical length should be at least $d\simeq200$ when
$\Delta_{2}/\Gamma=-5$. To achieve a reasonable signal (linear
transmission $T_{1}=1\%$) an optical density of $OD=3500$ is
needed. Importantly, however, we have shown that the presence of
bound states inside the nonlinear medium can be probed with
classical light, simply by examining higher-order correlation
functions in the output field, rather than sending in complicated
quantum inputs.

\subsection{Dissipative Regime $(\kappa=i|\kappa|)$}

In this section, we study the transport properties of the system
in the presence of nonlinear absorption, and calculate its effect
on the transmitted light and its correlation functions.

\begin{figure}[h]
\centering
\includegraphics[width=0.4\textwidth]{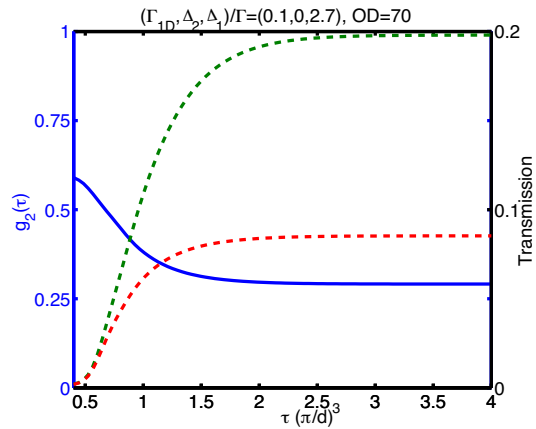}
\caption[Reaching the steady-state: absorptive case]{$g_{2}(\tau=0)$ reaches the steady-state after a time interval which
is set by the bandwidth of the system, one-photon state (green) is
partially transmitted while the two-photon state (red) is strongly
attenuated due the nonlinear absorption. This plot has been generated
for a system with $\Gamma_{1D}/\Gamma=10\%$ an OD=70.}
\label{fig:filter_absorption}
\end{figure}

\begin{figure}[h]
\centering
\includegraphics[width=0.4\textwidth]{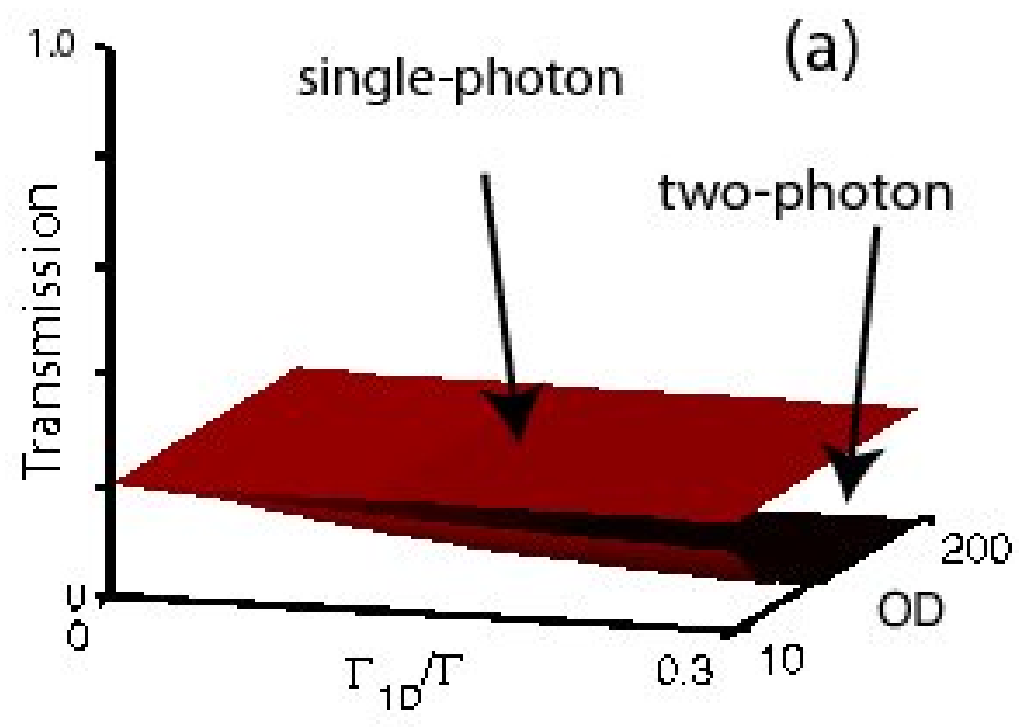}
\includegraphics[width=0.4\textwidth]{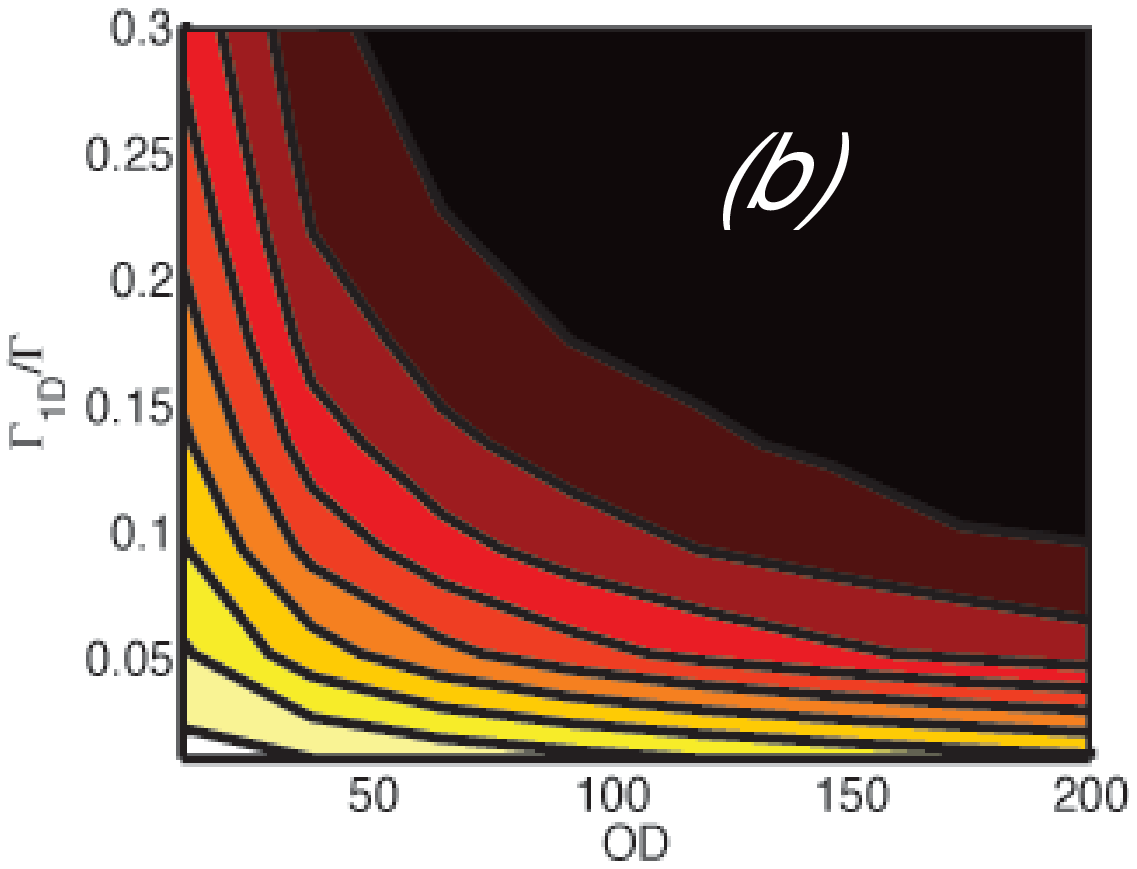}
 \caption[$g_2$ versus optical density and cooperativity: absorptive case]{(a) one-photon state is
partially transmitted ($T_1$) while the two-photon state transmission 
($T_2$) is suppressed due the nonlinear absorption. This suppression is more pronounced for higher optical density and cooperativity. (b) Correlation
function $g_2(\tau=0)$ of the transmitted light when the frequency is set to the single-photon transmission resonance with $T\simeq 20\%$ and $\frac{\Delta_2}{\Gamma}=0$.}
\label{fig:filter_scaling}
\end{figure}

A purely absorptive nonlinearity arises when the detuning
$\Delta_{2}$ is set to zero in our atomic system~(see
Fig.~\ref{fig:four_level}(a)). This nonlinear loss also leads to
anti-bunching in the transmitted field, as multi-photon components
become less likely to pass through the waveguide without being
absorbed. Linear absorption, on the other hand, affects
transmission of single- and multi-photon components equally. Fig.
\ref{fig:filter_absorption} and Fig.~\ref{fig:filter_scaling} show how two-photon and one-photon
states are transported differently in the nonlinear absorptive
system~(realistic linear losses are included in this calculation).


\begin{figure}[h]
\centering
\includegraphics[width=0.45\textwidth]{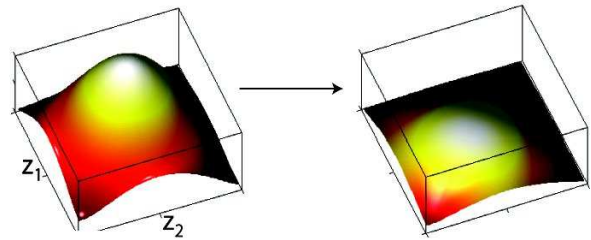}

\caption[Two-photon wave function suppression due to nonlinear absorption]{ In the presence of the nonlinear absorption ($\frac{\Delta_2}{\Gamma}=0$), the two-photon wave function is strongly suppressed comparing to the absence of the nonlinear absorption ($\Delta_2\gg\Gamma$). These plots has been generated for a system with
$\Gamma_{1D}/\Gamma=10\%$, $T_{1}\simeq20\%$ and OD=70.   }
\label{fig:quantum_dissipative}
\end{figure}

We note that the two-photon wavefunction is
attenuated due the nonlinear absorption, while it is not deformed,
as shown in Fig. \ref{fig:quantum_dissipative}. In
an experimental realization of such a system with
$\Gamma_{1D}/\Gamma=10\%$, an optical coherent length of
$d\simeq20$ is enough to yield a relatively large anti-bunching
($g_{2}<0.3$). In order to have high transmission ($T_{1}=20\%$)
for single photons an optical density of $OD \simeq 70$ is
required.

All of the physics related to the photon correlation function is
described again by product of the coherent optical length and the
nonlinearity coefficient $(|\kappa| d)$ (since the nonlinear absorption is equal to the nonlinear absorption coefficient times the length of the medium). However, for a fixed
optical density, since the nonlinear transition is on resonance,
the magnitude of the nonlinear coefficient $|\kappa|$ is enhanced
compared to the nonlinear dispersive case. We note that in the
presence of nonlinear absorption, one has to also consider the
effect of accompanied noise. However, the effect of noise for an ensemble of
many atoms which are driven by a weak laser field, is
negligible, and therefore, using the NLSE with a decay term is
sufficient and consistent. A rigorous demonstration of the validity of such approximation is the subject of further research.

\section{Conclusions\label{sec:Conclusion}}

We have developed a technique to study few-photon quantum dynamics
inside 1D nonlinear photonic system. This technique allows us to
study the system even in regimes where nonlinearities are
significant even at a few-photon level, where we find that the
behavior of the system deviates significantly from estimates based
on classical formalism. Specifically, when the system is driven by
classical light, the strong optical nonlinearity manifests itself
in the correlation functions of the outgoing transmitted light. In particular,
when the interaction between photons is effectively repulsive, the
suppression of multi-photon components results in anti-bunching of
the transmitted field and  the system acts as a single-photon switch. In the case of attractive
interaction, the system can exhibit either anti-bunching or
bunching, associated with the resonant
excitation of bound states of photons by the input field. These effects can be observed by probing statistics of photons transmitted through the nonlinear fiber.

\begin{acknowledgments}
We thank Anders S\o rensen, Victor Gurarie, Adilet Imambekov, and Shanhui Fan for
useful discussions. This work was partially supported by  NSF, NSF DMR-0705472, Swiss NSF, CUA, DARPA, Packard Foundation and AFOSR-MURI. DEC acknowledges support from the Gordon and Betty Moore Foundation through
Caltech's Center for the Physics of Information, and the National
Science Foundation under Grant No. PHY-0803371.

\end{acknowledgments}

\appendix

\section{EIT-Bandgap\label{sec:EIT-Bandgap}}

 In this appendix, we show that how in an EIT system, where the control field is a standing
wave, a band gap structure can be developed. In particular, we show the
presence of transmission resonances at the band gap edge by taking
into account the full expression for the atomic susceptibilities. We show
that at the band gap edge, we recover that same resonances that we presented in the main text for small detunings.

We consider a $\lambda$-level scheme, where a standing control field has coupled the forward- and backward-going probe together, similar to Fig.\ref{fig:four_level} without the nonlinear transition (c-d). Following \cite{Andre:Thesis}, we assume the noises to be negligible,
and therefore, the atomic equations of motion to the leading order
in $\mathcal{E_{\pm}}$ are

\begin{eqnarray}
\partial_{t}\hat{\sigma}_{ab}^+ & = & +(i\Delta_{1}-\Gamma/2)\hat{\sigma}_{ab}^++i\Omega \hat{\sigma}_{ac} +ig\sqrt{2\pi}\mathcal{E}_{+}\\
\partial_{t}\hat{\sigma}_{ab}^- & = & +(i\Delta_{1}-\Gamma/2)\hat{\sigma}_{ab}^-+i\Omega \hat{\sigma}_{ac} +ig\sqrt{2\pi}\mathcal{E}_{-}\\
\partial_{t}\hat{\sigma}_{ac} & = & -\gamma_{0}\hat{\sigma}_{ac}+i\Omega\hat{\sigma}_{ab}^++i\Omega\hat{\sigma}_{ab}^-\end{eqnarray}
and the evolution equation of the photonic fields are written as:

\begin{eqnarray}
(\partial_{t}+c\partial_{z})\mathcal{E}_{+} & = & i\Delta K\mathcal{E}_{+}+ig\sqrt{2\pi}n_0\hat{\sigma}_{ab}^+\\
(\partial_{t}-c\partial_{z})\mathcal{E}_{-} & = & i\Delta K\mathcal{E}_{-}+ig\sqrt{2\pi}n_0\hat{\sigma}_{ab}^-.\end{eqnarray}

The wave-vector mismatch can be ignored by including a small shift in the two-photon detuning. By taking the Fourier transform of the
atomic equation of motion, one can solve for atomic polarization
and obtain the self- and cross-susceptibilities. We can define a
unit length based on the absorption length
$L_{abs}=\frac{c\Gamma}{2\pi g^{2}n_0}$, and write the field equation as:

\begin{eqnarray}
\partial_{\tilde{z}}\mathcal{E}_{+} & = & i\tilde{\Delta}_3\mathcal{E}_{+}+i\chi_{s}(\delta)\mathcal{E}_{+}+i\chi_{c}(\delta)\mathcal{E}_{-}\label{eq:coupled-mode_full}\\
-\partial_{\tilde{z}}\mathcal{E}_{-} & = & i\tilde{\Delta}_3\mathcal{E}_{-}+i\chi_{s}(\delta)\mathcal{E}_{-}+i\chi_{c}(\delta)\mathcal{E}_{+}\nonumber \end{eqnarray}
where the self- and cross susceptibilities and the detuning are given
by:

\begin{eqnarray}
\chi_{s}(\delta) & = & i\frac{\Gamma}{\Gamma'}\frac{\Gamma'\Gamma_{0}+\Omega^{2}}{\Gamma'\Gamma_{0}+2\Omega^{2}}\\
\chi_{c}(\delta) & = & -i\frac{\Gamma}{\Gamma'}\frac{\Omega^{2}}{\Gamma'\Gamma_{0}+2\Omega^{2}}\\
\tilde{\Delta}_{3} & = & \frac{\Delta_{3}}{\Gamma}\frac{\Gamma^{2}}{2\pi g^{2}n_0}=\frac{\delta}{\eta}\frac{\Gamma}{|\Delta_1|}\end{eqnarray}
where $\Gamma'=\Gamma/2-i\Delta_{1}-i\Delta_{3}$ , $\Gamma_{0}=\gamma_{0}-i\Delta_{3}$
and $\Delta_{3}$ is the two-photon detuning of the probe from the pump field which is related to the
dimensionless two-photon detuning in the main text ($\Delta_{3}=2\frac{\Omega^{2}}{\Delta_{1}}\delta\ll \Delta_1$).
We note that in most cases, $\tilde{\Delta}_{3}$ is very small for
slow group velocities ($\frac{\Gamma^{2}}{2\pi g^{2}n_0}=(\frac{\Gamma}{\Omega})^{2}\frac{\Omega^{2}}{2\pi g^{2}n_0}\ll1$),
and therefore the corresponding term can be neglected for simplicity.

In order to obtain transmission and reflection coefficient, one should solve the couple mode equations Eq.(\ref{eq:coupled-mode_full}) with proper boundary conditions. Therefore, we consider a system which is driven with a weak coherent field
at $(z=0)$. Therefore, the boundary conditions can be set to,

\begin{eqnarray}
\mathcal{E}_{+}(z=0) & = & \mathcal{E}_0\\
\mathcal{E}_{-}(z=d) & = & 0.\end{eqnarray}

\begin{figure}
\centering
\includegraphics[width=0.3\textwidth]{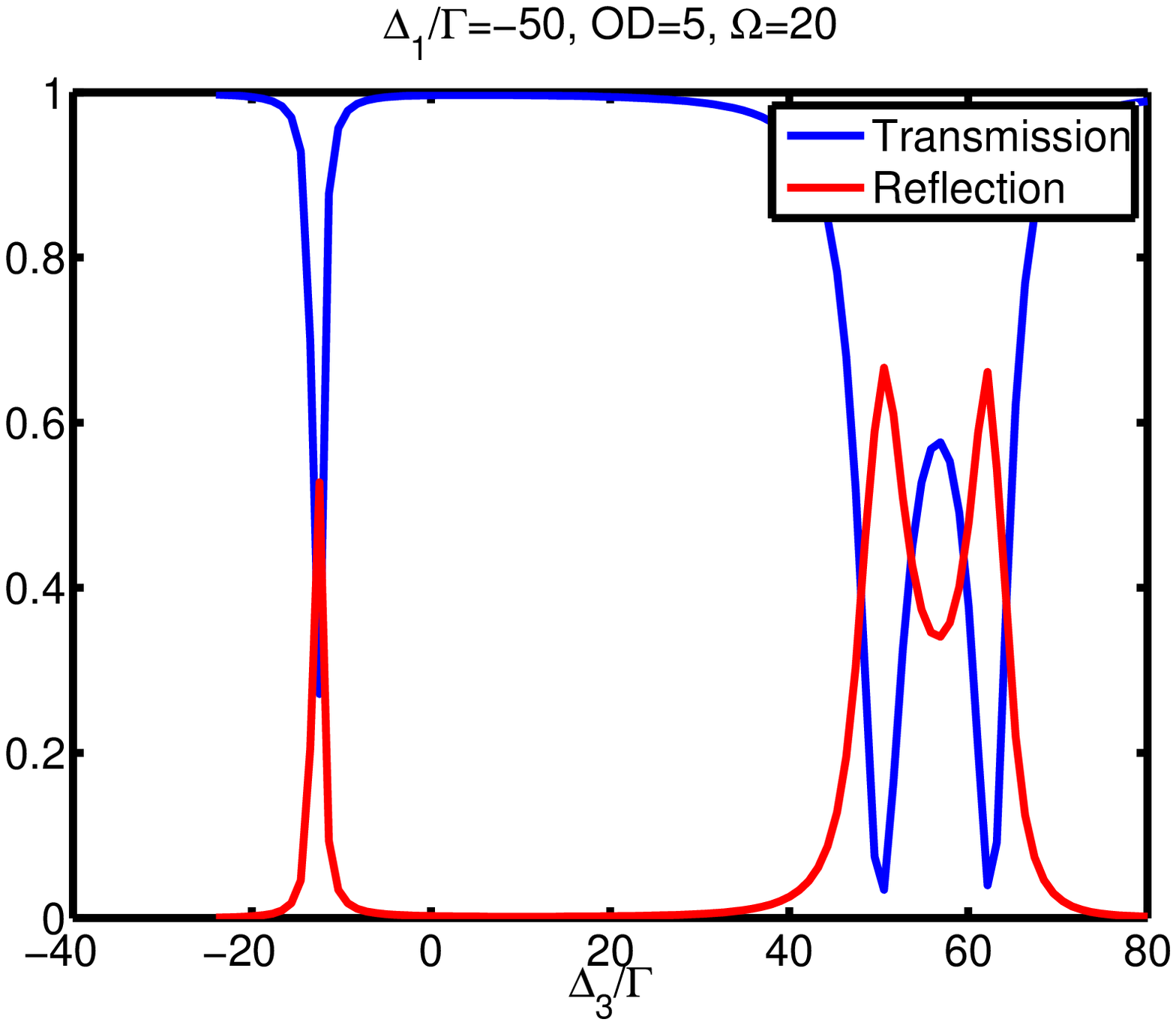}
\centering
\includegraphics[width=0.3\textwidth]{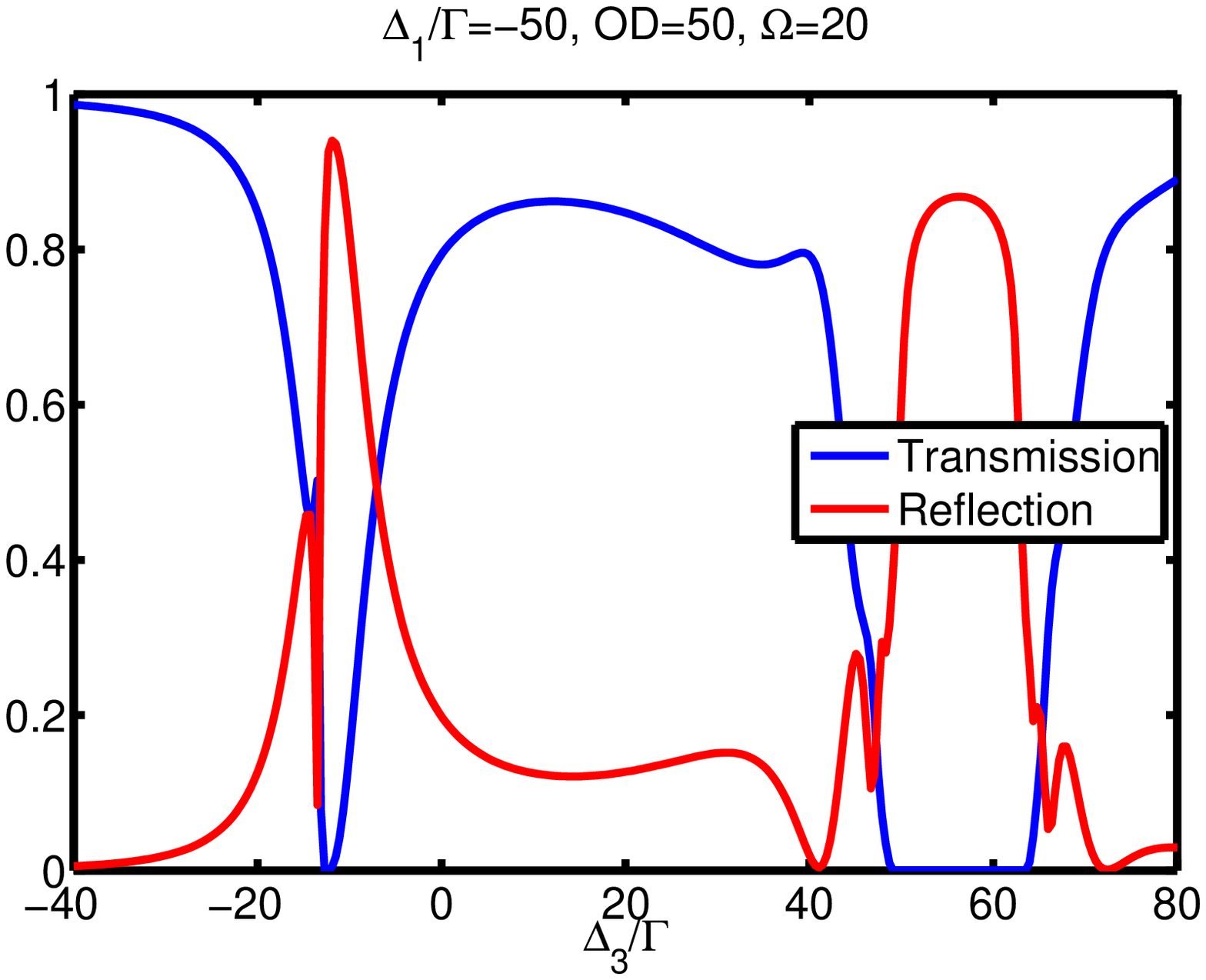}
\includegraphics[width=0.3\textwidth]{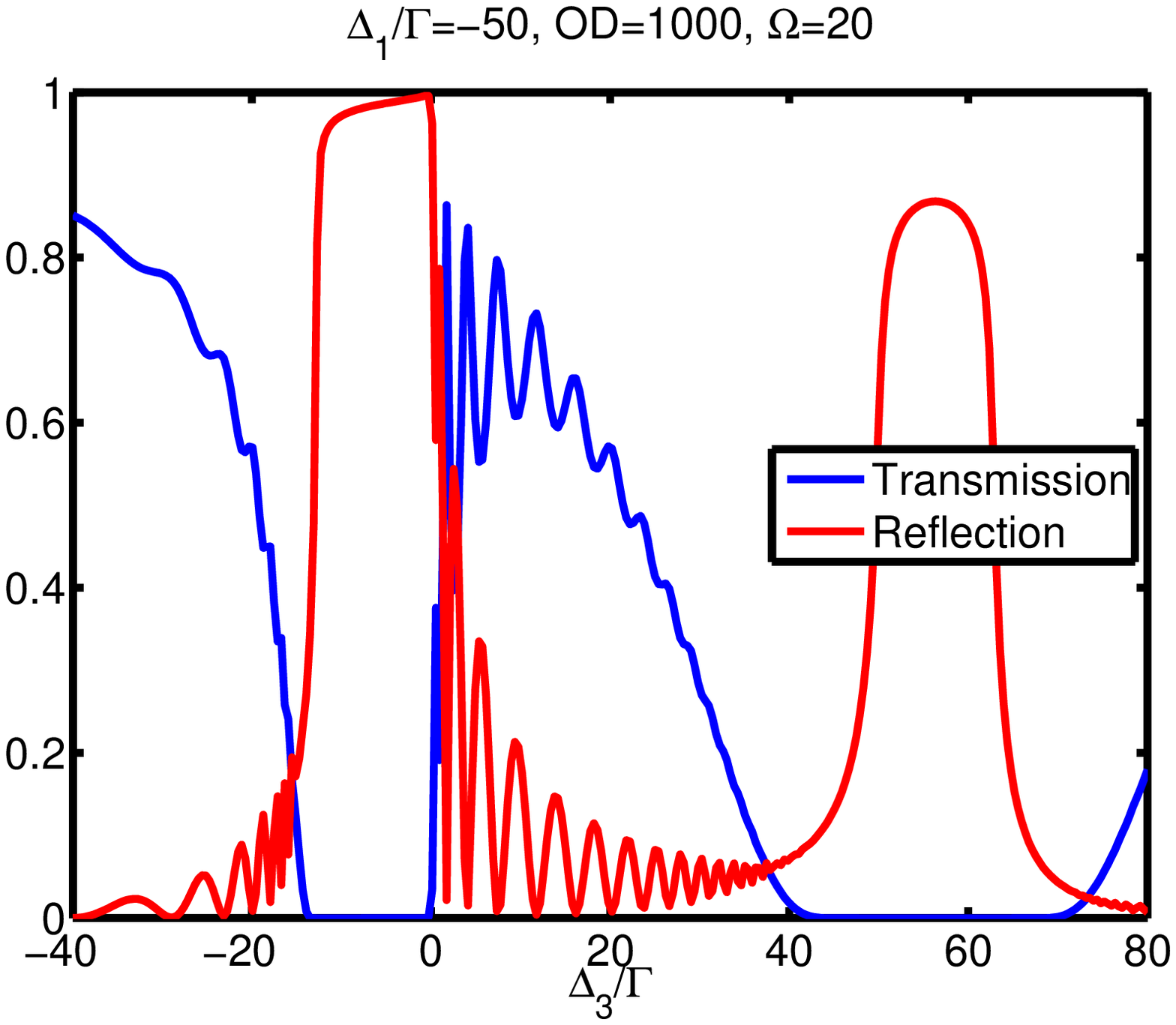}\caption[Band gap formation for different optical densities ]{By increasing OD, the band gap structure becomes more pronounced\label{fig:vary_OD}}

\end{figure}

We evaluate the transmission coefficient $(\mathcal{E}_{+}(z=d)/\mathcal{E}_0)$,
and the reflection coefficient $(\mathcal{E}_{-}(z=0)/\mathcal{E}_0)$ by numerical
methods using BVP5C in Matlab. In particular, we are interested in
the Raman regime, in other words the detuning is very large $|\Delta_{1}|\gg\Gamma$
and also we assume $\Delta_{1}<0$. First, we consider the case where
the EIT width is smaller than the one-photon detuning, i.e. $\Omega\ll|\Delta_{1}|$.
Fig.\ref{fig:vary_OD} shows the reflectivity and transmittivity of
the system for different optical densities. In the regime with low
optical density, the spectrum corresponds to a shifted Raman transition
at $\Delta_{3}\simeq2\frac{\Omega^{2}}{\Delta_{1}}$ and an EIT window
around $\Delta_{3}\simeq-\Delta_{1}$. In higher optical densities,
the system develops a band gap for $-\Omega\leq\Delta_{3}\leq0$. Fig.\ref{fig:vary_OD}
shows that in media with higher optical density, the band gap becomes
more prominent.

\begin{figure}[h]
\centering
\includegraphics[width=0.4\textwidth]{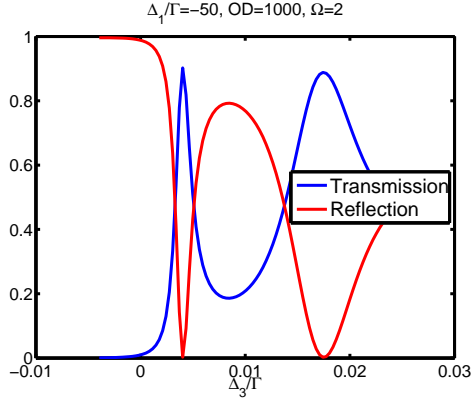}

\caption[Band gap edge]{Transmission resonances at the edge of the band gap\label{fig:edge_bandgap}}

\end{figure}

As we discussed in the main text, we are interested in the band gap
edge where the transmission peaks are present and the system acts
like an effective cavity. Fig.\ref{fig:edge_bandgap} shows a close-up of the
transmittivity and reflectivity spectrum in Fig.\ref{fig:vary_OD}(c) at the band gap edge. We can observe that several resonances occur
at the edge due to the finite size of the system. By positioning at
the one of the transmission peaks, the system behaves as an effective
cavity, where the decay rate of the cavity will be given by the width
of the transmission peak. Therefore, the present results, including
the full susceptibilities of the system, is consistent with the model
presented earlier where we had approximated the system to be around
$\Delta_{3}=0$.

\begin{figure}[h]
\centering
\includegraphics[width=0.4\textwidth]{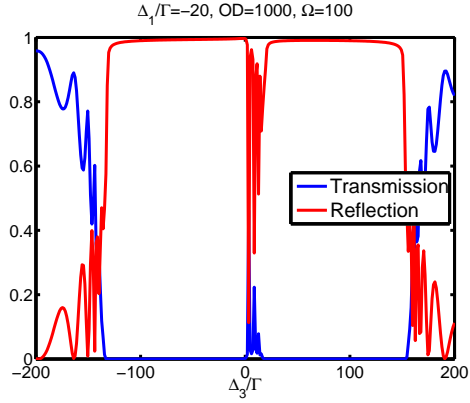}

\caption[Band gap structure for strong control fields]{Band gap structure for strong control fields $(\Omega>\Delta_{1}$)\label{fig:EIT_bandgap}}

\end{figure}

We add that alternatively, one can assume a strong control field so
that the EIT windows would be smaller than the one-photon detuning
$\Omega>|\Delta_{1}|$. Similar to the previous case $\Omega<|\Delta_{1}|$,
the system develops a band gap. As shown in Fig.\ref{fig:EIT_bandgap},
the band gap is formed between $-\Omega\lesssim\Delta_{3}\lesssim\Omega$,
similar to modulated EIT with AC-stark shift as discussed in Ref.\cite{Andre:bandgap}.

\section{Numerical Methods \label{sec:numerics}}

\label{sec:numerics}

In this section, we describe the numerical methods that have been
used to simulate the evolution of the photonic quantum state and
the related correlation functions, in the limit where we truncate
the Hilbert space to two photons or less. The partial differential
equations for the one-photon and two-photon
wave functions~(\ref{eq:two_photon}, \ref{eq:one_photon}) are turned
into difference equations by discretizing space and time, and are
evolved forward in time using the Du Fort-Frankel
scheme~\cite{Strikwerda:2004}. This algorithm is is explicit in
time -- \textit{i.e.}, the next time step function is explicitly
given by the past time function -- and is also unconditionally
stable. We note that the system under investigation is open and it
is driven out of equilibrium, therefore, conventional analytical
methods for approaching the NLSE such as Bethe ansatz or quantum
inverse scattering~\cite{korepin:1993} are not applicable here.

The one-photon wave function can be easily integrated and solved
analytically. However, we describe how to obtain the one-photon
wave function numerically and then generalize this technique to
obtain the two-photon wave function. First, we mesh space and time
and reduce the differential equations to a difference equation. If
we choose the time step $k$ and the space step $h$, the discretized time and space will be $x=z/h$ and $s=t/k$ and the system length $d=Nh$.  Then following the Du
Fort-Frankel scheme~\cite{Strikwerda:2004}, the evolution equation
takes the form:
\begin{eqnarray}
\frac{\theta(x,s+1)-\theta(x,s-1)}{2k}&=&\frac{i}{2mh^{2}}[\theta(x+1,s)+\theta(x-1,s)\nonumber\\
&-&\theta(x,s+1)-\theta(x,s-1)]\end{eqnarray}
where the position take all values inside the boundary ($2\leq
x\leq N-1$). By
rearranging the above equation, the explicit form of the equation
can be obtained
\begin{eqnarray}
(&1&+~\frac{ik}{mh^{2}})\theta(x,s+1)=\theta(x,s-1)\\
&+&\frac{ik}{mh^{2}}\left[\theta(x+1,t)+\theta(x-1,t)-\theta(x,s-1)\right]\nonumber\end{eqnarray}
Therefore inside the boundaries, the wave function at time $s+1$
can be obtained knowing the wave function at time $s$ and $s-1$.
The boundary condition at $z=0$ -i.e. $x=1$, will be given by
\begin{eqnarray}
\alpha&=&\frac{\theta(1,s+1)+\theta(2,s+1)}{2}\\
&-&i\frac{\theta(2,t+1)-\theta(1,s+1)}{2mh}\nonumber\end{eqnarray}
 Or equivalently, \begin{equation}
\theta(1,s+1)=\frac{\alpha+(-\frac{1}{2}+\frac{i}{2mh})\theta(2,s+1)}{\frac{1}{2}+\frac{i}{2mh}}\end{equation}
 and similarly for the boundary condition at $z=d$ -i.e. $x=N$,
we have \begin{equation}
\theta(N,s+1)=\frac{(-\frac{1}{2}+\frac{i}{2mh})\theta(N-1,s+1)}{\frac{1}{2}+\frac{i}{2mh}}.\end{equation}
 Therefore, by having the above boundary conditions and the initial
condition $\theta(x,s=1)=0$, the wave function can be calculated
at any time inside the boundaries ($2\leq x\leq N-1$). The order
of accuracy of the Du Fort-Frankel scheme is given by
$O(h^{2})+O(k^{2})+O(k^{2}h^{-2})$ and it is consistent as $k/h$
tends to zero \cite{Strikwerda:2004}.

Similarly, we can write a difference equation for the two-photon
wave function. The $\delta$-interaction can be approximated by a
Gaussian distribution. The space domain is meshed so that $\Delta
z_{1}=\Delta z_{2}=h$. The evolution equation for the two-photon
wave function reads
\begin{eqnarray}
&~&\left(1+\frac{2ik}{mh^{2}}\right)\phi(x,y,s+1) = \\
&~&\frac{ik}{mh^{2}}\left[\phi(x+1,y,s)+\phi(x-1,y,s)\right]\nonumber\\
&~&+ \frac{ik}{mh^{2}}\left[\phi(x,y+1,s)+\phi(x,y-1,s)\right]\nonumber\\
&~&+ 2\frac{ik}{mh^{2}}\phi(x,y,s-1)+\phi(x,y,s-1)\nonumber\\
&~&- 2 k  \frac{2i\kappa}{\sigma\sqrt{2\pi}}{\rm {Exp}\left(-\frac{(x-y)^{2}}{2\sigma^{2}}\right)\phi(x,y,s)}\nonumber\end{eqnarray}
 The boundary condition at $z=0$ will be given by 
 \begin{eqnarray}
\frac{\alpha}{2}\theta(y,s+1)&=&\frac{\phi(1,y,s+1)+\phi(2,y,s+1)}{2}\\
&-&i\frac{\phi(2,y,s+1)-\phi(1,y,s+1)}{2mh}\nonumber,\end{eqnarray}
where $\sigma$ is the length scale characterizing the distance of
the two-photon interaction. Approximating the delta-function with a Gaussian is valid if $\sigma\ll d$. On the other hand, we should have $h\ll \sigma$ so that the Gaussian function would be smooth. Or
equivalently,
\begin{equation}
\phi(1,y,s+1)=\frac{\frac{1}{2}\alpha\theta(y,s+1)+(-\frac{1}{2}+\frac{i}{2mh})\phi(2,y,s+1)}{\frac{1}{2}+\frac{i}{2mh}}\end{equation}
 and similarly for the boundary condition at $z=d$, we have \begin{eqnarray}
0&=&\frac{\phi(N,y,s+1)+\phi(N-1,y,s+1)}{2}\\
&+&i\frac{\phi(N,y,s+1)-\phi(N-1,y,s+1)}{2mh}\nonumber\end{eqnarray}
 which gives \begin{equation}
\phi(N,y,s+1)=\frac{(-\frac{1}{2}+\frac{i}{2mh})\phi(N-1,y,s+1)}{\frac{1}{2}+\frac{i}{2mh}}.\end{equation}
Once the wave function is known at any point in time and space, we
can evaluate the correlation functions. In particular, the
two-photon correlation function $g_{2}(d,\tau=0)$ is given by
Eq.\,(\ref{eq:g2_0}), where the first and the second derivatives
at anytime are given by the following expressions,

\begin{eqnarray}
\partial^{(1)}\phi(d,d) & =&  \frac{1}{2mh}[  \phi(N,N)  -\phi(N-1,N)]\\
\partial^{(1)}\partial^{(2)}\phi(d,d) & =&  \frac{1}{4m^{2}h^{2}}[  \phi(N,N)  -\phi(N-1,N)\nonumber\\
&- & \phi(N,N-1)  +\phi(N-1,N-1)].\nonumber \end{eqnarray}
 Note that in evaluation of $g_{2}(\tau)$, once the first photon is
detected the two-photon wave function collapses to zero. This
seems to be contradictory with the driven boundary condition
Eq.(\ref{eq:boundary_condition}) where the two-photon state at the
boundaries is proportional to the one-photon wave function which
is not zero. This apparent inconsistency occurs because we have
neglected higher number photon states in our truncation. However,
this inconsistency only leads to higher order corrections to
$g_{2}(\tau)$ in the input field amplitude $\alpha$, which is
assumed to be weak~($\alpha{\ll}1$).

\end{document}